\documentclass[APA,LATO1COL]{WileyNJD-v2}
\usepackage{amsfonts}
\usepackage{animate}
\usepackage{epsfig, epsf, graphicx}

\articletype{Article Type}%

\received{26 April 2019}
\revised{6 June 2019}
\accepted{6 June 2019}

\begin{document}

\title{Trajectory Functional Boxplots}

\author[1]{Zonghui Yao}

\author[2]{Wenlin Dai*}

\author[1]{Marc G. Genton}

\authormark{ZONGHUI YAO \textsc{et al.}}

\address[1]{\orgdiv{Statistics Program}, \orgname{King Abdullah University of Science and Technology}, \orgaddress{\state{Thuwal 23955--6900}, \country{Saudi Arabia}}}

\address[2]{\orgdiv{Institute of Statistics and Big Data}, \orgname{Renmin University of China}, \orgaddress{\state{Beijing}, \country{China}}}

\corres{*Wenlin Dai, Institute of Statistics and Big Data, Renmin University of China, Beijing, China.\\
\email{wenlin.dai@ruc.edu.cn}}

\abstract[Summary]{With the development of data-monitoring techniques in various fields of science, multivariate functional data are often observed. Consequently, an increasing number of methods have appeared to extend the general summary statistics of multivariate functional data. However, trajectory functional data, as an important sub-type, have not been studied very well. This article proposes two informative exploratory tools, the trajectory functional boxplot, and the modified simplicial band depth (MSBD) versus Wiggliness of Directional Outlyingness (WO) plot, to visualize the centrality of trajectory functional data. The newly defined WO index effectively measures the shape variation of curves and hence serves as a detector for shape outliers; additionally, MSBD provides a center-outward ranking and works as a detector for magnitude outliers. Using these two measures, the functional boxplot of the trajectory reveals center-outward patterns and potential outliers using the raw curves, whereas the MSBD-WO plot illustrates such patterns and outliers in a space spanned by MSBD and WO. The proposed methods are validated on hurricane path data and migration trace data recorded from two types of birds.}


\maketitle

\section{Introduction}\label{s1}
\noindent
Due to the rapid progress in data-monitoring techniques and the Internet, the volume of data has experienced explosive growth. 
Functional data are commonly recorded among various fields, including, but not limited to, medical imaging, meteorology, biology, and engineering. 
Examples include temperature and precipitation records at weather stations, hand-writing data in different languages, and absorption curves of some medical ingredients. 
Responses at points of observation are categorized as univariate or multivariate functional data. 
Functional data analysis has attracted great attention over the last two decades \citep{ramsayfunctional,ferraty2006nonparametric,horvath2012inference}; we refer the readers to \citet{wang2016functional} for a recent review. 
Most research focuses on the univariate cases, leaving the multivariate cases less explored.

\begin{figure}[t!]
\centering
\includegraphics[width=.328\textwidth]{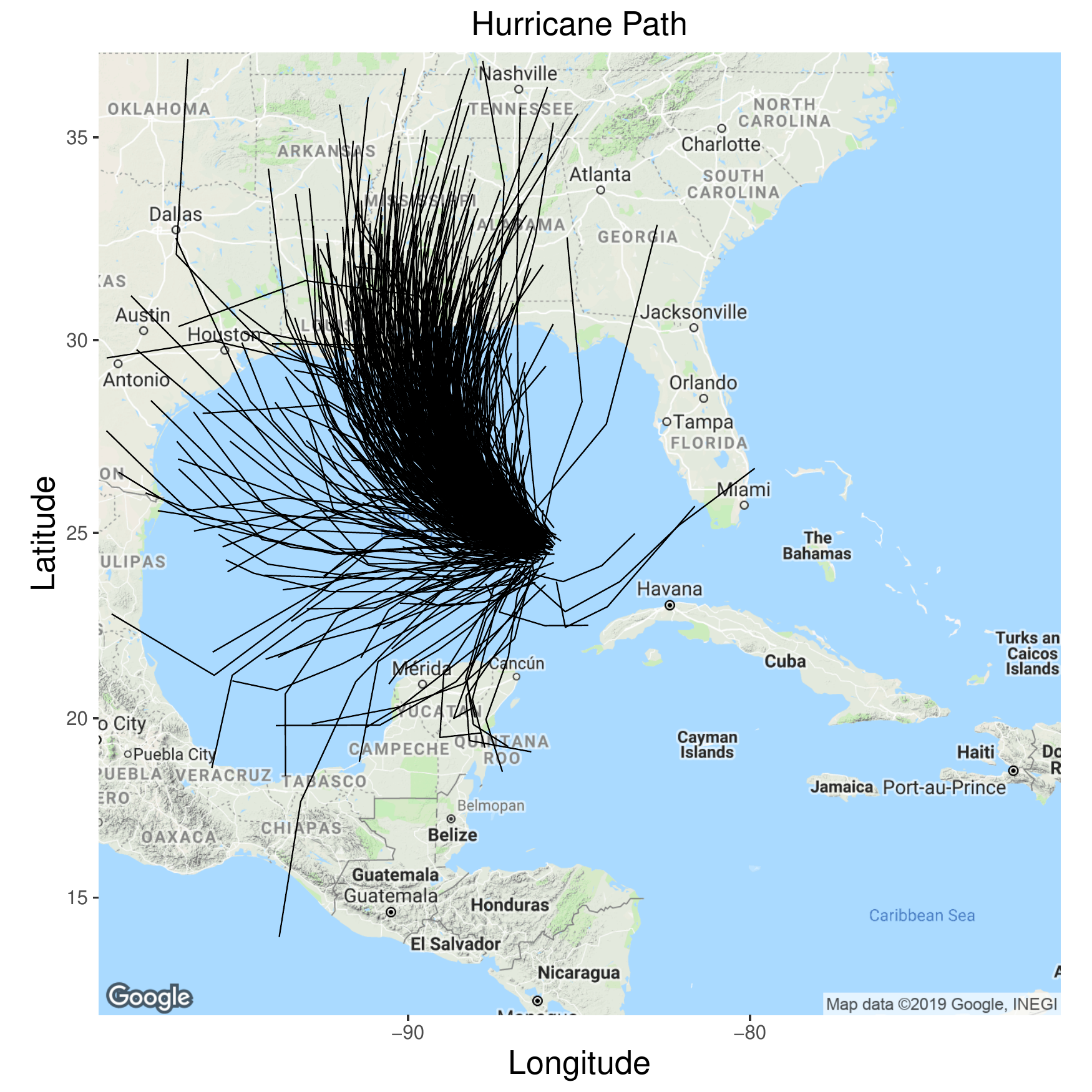}
\includegraphics[width=.328\textwidth]{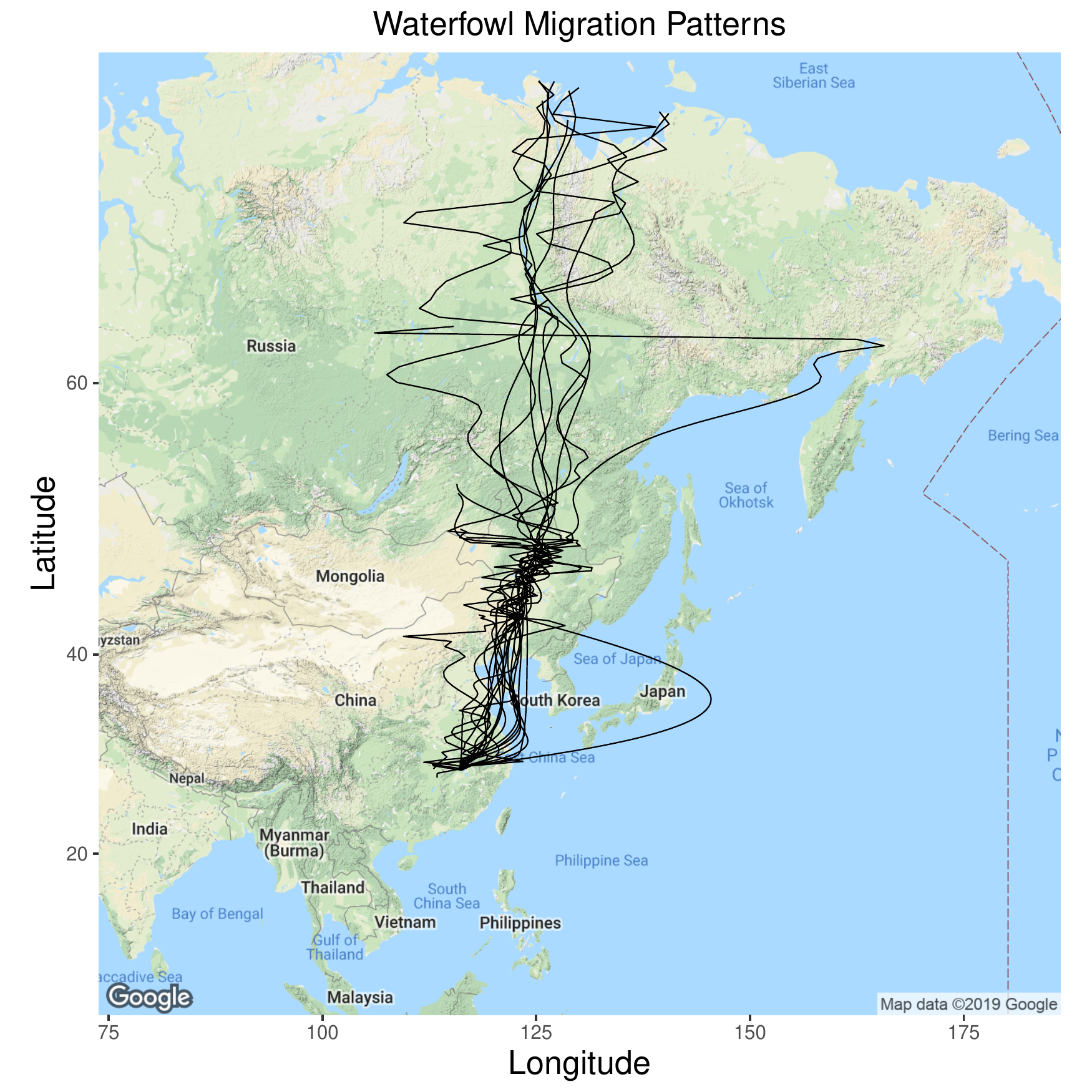}
\includegraphics[width=.328\textwidth]{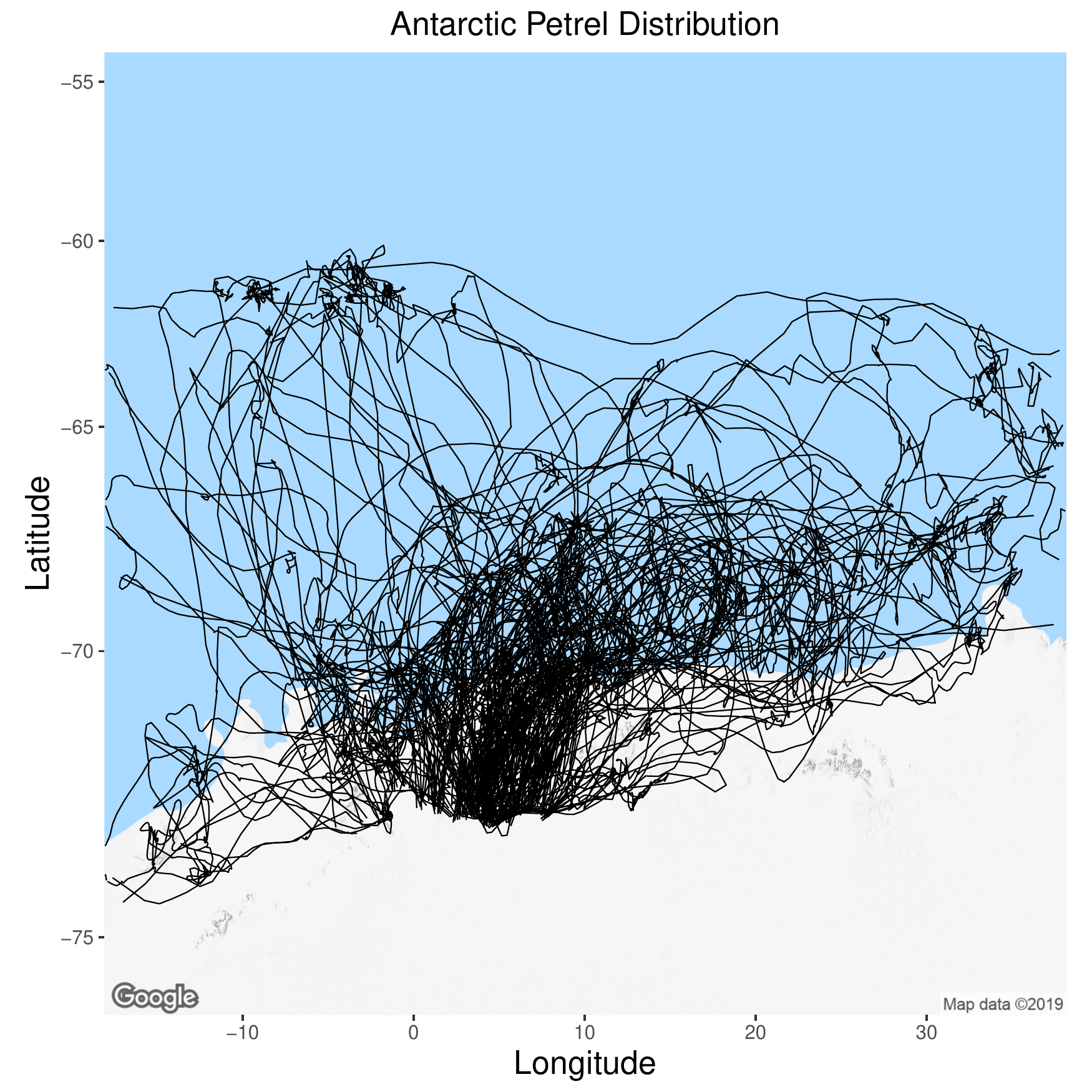}
(a)\hspace{5.3cm}(b)\hspace{5.3cm}(c)
\caption{Trajectory functional data: (a) hurricane path trajectories; (b) waterfowl migration path trajectories; (c) petrel migration trajectories.}
\label{1}
\end{figure}

Here, we focus on trajectory data, an important type of multivariate functional data. Trajectory data usually record the positions of objects during a specific time window and commonly appear in many important research areas. We provide three examples in Figure {\color{blue}{\ref{1}}} that include the hurricane paths from a predictive model \citep{cox2013visualizing} and the migration paths of two types of birds \citep{descamps2016sea,si2018spring}. We propose to develop some tools for exploratory analysis, specifically for this type of data.  

During the boom of functional data analysis, many summary statistics and inference techniques have been generalized from traditional to functional data. However, trajectory data have not been sufficiently investigated, and the corresponding ranking methods, outlier detections, and visualizations remain open questions. 
Most existing exploratory analysis methods for functional data are based on the concept of statistical depth, which is initially a potent tool to rank multivariate data but also does well in describing the centrality for functional data. Several depth notions have been proposed to rank multivariate functional data, e.g., weighted modified band depth (WMBD; \citeauthor{ieva2013depth}, \citeyear{ieva2013depth}), simplicial band depth and modified simplicial band depth (SBD and MSBD; \citeauthor{lopez2014simplicial}, \citeyear{lopez2014simplicial}); they are the prevailing methods to give a plausible center-outward sequence. \cite{dai2018directional} introduced the directional outlyingness for detecting outliers from multivariate functional data.

Outlier detection is another crucial step in the analysis of data. 
The well-known types of functional outliers include persistent outliers, isolated outliers, magnitude outliers, and shape outliers (\citeauthor{hubert2015multivariate}, \citeyear{hubert2015multivariate}). The first three types of outliers can be handled by the simplicial band depth. However, shape outlier detection is a more challenging task.
 Shape outliers are defined as trajectories exhibiting a different shape from the rest of the sample. The outliergram (\citeauthor{arribas2014shape}, \citeyear{arribas2014shape}) is one choice for shape outlier detection, based on the modified epigraph index and the modified band depth, but they only show its capacity in the univariate case. \cite{dai2018directional} combined the magnitude and shape outlyingness through forming vectors of the mean of directional outlyingness (MO) and variance of directional outlyingness (VO), then calculated their Robust Mahalanobis Distance (RMD) with the minimum covariance determinant estimator of \cite{rousseeuw1985multivariate}. They defined the outliers as those for which RMD values are beyond a specific threshold. However, this method cannot detect the two types of outliers, shape and magnitude, separately. Thus, it leads to large false detection rates. 

Visualization tools are  commonly used to illustrate the properties of the analyzed data. For functional data, various tools have been developed, such as functional bagplots and functional highest density region plots (\citeauthor{hyndman2010rainbow}, \citeyear{hyndman2010rainbow}), functional boxplots  (\citeauthor{sun2011functional}, \citeyear{sun2011functional}), and surface boxplots (\citeauthor{genton2014surface}, \citeyear{genton2014surface}). These plots give a good description of the functional data and show each curve directly with different labels. Another type of plots is based on the magnitude versus shape index of each curve, showing the centrality of data by scatterplots. Outliergrams \citep{arribas2014shape}, functional outlier maps (\citeauthor{rousseeuw2018measure}, \citeyear{rousseeuw2018measure}), and magnitude-shape plots (\citeauthor{dai2018multivariate}, \citeyear{dai2018multivariate}) are some examples. Yet, a good visualization tool for trajectory functional data is lacking.

In this paper, we propose two visualization tools for trajectory functional data analysis. Specifically, we develop the ``Wiggliness of Directional Outlyingness" (WO), which performs very well in detecting shape outliers in trajectory functional data. Based on the results, we first construct a trajectory functional boxplot, that visualizes the raw curves with different percentage bands and outliers; we then provide another scatterplot, the MSBD-WO plot, presenting the magnitude and shape properties for each curve.

The remaining of the paper is organized as follows: Section \ref{s2} introduces trajectory functional data and commonly used methods for curve ranking and outlier detection. Section \ref{s3} provides the two visualization tools constructed using a new measure of centrality defined especially for trajectory functional data. Section \ref{s4} compares the performance of the proposed procedures with several outlier detection methods in a series of simulation studies, and Section \ref{s5} presents three data applications of the proposed tools. A conclusion is provided in Section \ref{s6}.

\section{Trajectory Functional Data}  \label{s2}

\quad Trajectory functional data naturally appear in many situations, such as weather forecasting, ecological studies, and handwriting inputs. They are special forms of multivariate functional data. The main difference is that, instead of visualizing the data along time, the data are mapped in a sub-space by removing the time axis. Figure {\color{blue}{\ref{1}(a)}} shows classical hurricane trajectory data that record the locations of hurricanes with time. Instead of showing the graph in 3D, we plot the trajectories on a 2D map. We can treat trajectory functional data as a $p$-dimensional stochastic process $\textbf{X}(t)$, where $t$ is defined on a compact interval $\mathcal{I}$. In the hurricane path example, $p=2$. Often, the trajectories of a sample share approximately the same starting and${/}$or ending points; otherwise, an alignment step should be implemented before analyzing the data.

\subsection{Multivariate Curve Ranking}

A natural way to rank these trajectory functional data is to use a depth notion for multivariate functional data to make a center-outward ordering for the curves that provides a robust description of the data structure. Here, we consider the following two tools: the simplicial band depth (SBD) (\citeauthor{lopez2014simplicial}, \citeyear{lopez2014simplicial})  and the directional outlyingness (\citeauthor{dai2018directional}, \citeyear{dai2018directional}) to perform the ranking. Other possible methods for ranking multivariate functional data include multivariate functional halfspace depth \citep{claeskens2014multivariate}  and high-order integrated or infimal depth \citep{nagy2017depth}.

%
%
\subsubsection{Simplicial Band Depth} 
The simplicial band depth (SBD) (\citeauthor{lopez2014simplicial}, \citeyear{lopez2014simplicial}) is defined as
\begin{equation*}
{\rm SBD}({\textbf x},P_{\textbf X})=P\{{\textbf x}(t) \in {\rm simplex}\{{\textbf X}_1(t),\ldots,{\textbf X}_{p+1}(t)\}, \forall t \in \mathcal{I}\},
\end{equation*}
where we use a random ${\rm simplex}\{\textbf{X}_1(t),\ldots,\textbf{X}_{p+1}(t)\}$ in $\mathbb{R}^p$ defined by $\textbf{X}_1(t),\ldots,\textbf{X}_{p+1}(t)$. It measures the probability that the random regions in $\mathbb{R}^{p+1}$ decided by random simplices at time $t$ contain ${\textbf{x}}(t)$.

Because it is usually not likely for a curve to be completely incorporated in a simplex, \cite{lopez2014simplicial} relaxed the strict containment requirement, and formed a modified simplicial band depth (MSBD) as 
\begin{equation*}
{\rm MSBD}(\textbf{x},P_{\textbf{X}})={\rm E}(\lambda[t\in \mathcal{I},~ \mbox{s.t.} ~\textbf{x}(t)\in {\rm simplex}\{\textbf{X}_1(t),\ldots,\textbf{X}_{p+1}(t)\}]),
\end{equation*}
where $\lambda$ is the Lebesgue measure on $\mathcal{I}$ divided by the length of the interval $\mathcal{I}$. Obviously, this depth measures the time period during which the trajectory of $\textbf{x}(t)$ is incorporated in the simplices determined by $\textbf{X}_1(t),\ldots,\textbf{X}_{p+1}(t)$.

\subsubsection{Directional Outlyingness} 
Let $\textbf{X}(t)$ be a $p$-dimensional function defined on a domain $\mathcal{I}$. We define $d(\textbf{X}(t),F_{\textbf{X}(t)})$ as a depth function for $\textbf{X}(t)$ with respect to $F_{\textbf{X}(t)}$ which denotes the distribution of a random variable, and $o(\textbf{X}(t),F_{\textbf{X}(t)})$ as the corresponding outlyingness of $\textbf{X}(t)$, with respect to $F_{\textbf{X}(t)}$.

\quad In order to capture the shape as well as magnitude outliers, \cite{dai2018directional} introduced the following definition for directional outlyingness:
\begin{equation*} 
\textbf{O}(\textbf{X}(t), F_{\textbf{X}(t)}) =o(\textbf{X}(t), F_{\textbf{X}(t)}) \cdot \textbf{v}(t)=\{ 1/d(\textbf{X}(t),F_{\textbf{X}(t)})-1\} \cdot \textbf{v}(t),
\end{equation*}
where $\textbf{v}(t)$ is the unit vector pointing from the median of $F_{\textbf{X}(t)}$ to $\textbf{X}(t)$, $\textbf{v}(t)=\{\textbf{X}(t)-\textbf{Z}(t)\}/\|\textbf{X}(t)-\textbf{Z}(t)\|_2$, and $\textbf{Z}(t)$ stands for the median of the distribution $F_{\textbf{X}(t)}$. \cite{dai2018directional} suggested to use distance-based depths, e.g., Mahalanobis depth or projection depth \citep{zuo2000general}, to construct the directional outlyingness.

\cite{dai2018directional} defined two indices that measure the outlyingness of functional data: the mean of directional outlyingness ($\textbf{MO}$) and the variation of directional outlyingness ($VO$). In actual situations, we have only a finite set of time points. Therefore, ${\rm \textbf{MO}}_{T_k,n}(\textbf{X},F_{\textbf{X},n})$ and $VO_{T_k,n}(\textbf{X},F_{\textbf{X},n})$ are the measures used in real applications where $T_k=\{t_1,t_2,\ldots,t_k\}$.

\subsection{Outlier Detection} \label{sim}
When the underlying dataset is possibly contaminated, the detection of outliers becomes an important step of exploratory data analysis. For functional data, the existing outlier detection rules consist of three different subtypes: discarding a prefixed proportion of data with respect to the depth values \citep{fraiman2001trimmed}, using graphical tools based on the raw curves \citep{hyndman2010rainbow,sun2011functional,xie2017geometric}, and approximating the distribution of the depth (or its transformation) values \citep{rousseeuw2018measure,dai2018directional}. We use two of them that belong to the last two categories, respectively. 
\vspace{-.05cm}

\subsubsection{Simplicial Band Depth Criteria} 
The empirical rules of cutoff value are formed by a constant factor $F^*$ times the height of the 50\% central region ranked by the depth, where, usually, $F^*=1.5$ based on the simulation study conducted by \cite{sun2011functional,sun2012adjusted}. The definition of outliers under MSBD criteria identifies curves that cross the threshold.

\subsubsection{Robust Mahalanobis Distance Criteria}

Besides setting a cutoff value according to the functional depth distribution, Dai and Genton (\citeyear{dai2018directional}) showed that the distribution of $\textbf{Y}_{k,n}=(\textbf{MO}^\top_{T_k,n},{\rm VO}_{T_k,n})^\top$ could be asymptotically-approximated by a $p+1$ dimensional Gaussian distribution, if $\textbf{X}(t)$ was generated from a $p$-dimensional stationary Gaussian process.
They used the robust square Mahalanobis distance:
\begin{equation*}{\rm RMD}^2(\textbf{Y}_{k,n},\bar{\textbf{Y}}_{k,n,J})=(\textbf{Y}_{k,n}-\bar{\textbf{Y}}_{k,n,J})^\top\textbf{S}_{k,n,J}^{-1}(\textbf{Y}_{k,n}-\bar{\textbf{Y}}_{k,n,J}),
\end{equation*}
where $J$ is a group containing $h$ points that minimize the determinant of the corresponding covariance matrix. Here, $\bar{\textbf{Y}}_{k,n,J}=h^{-1}\sum_{i\in J}\textbf{Y}_{k,n,i}$ and $\textbf{S}_{k,n,J}=h^{-1}\sum_{i\in J}(\textbf{Y}_{k,n,i}-\bar{\textbf{Y}}_{k,n,J})(\textbf{Y}_{k,n,i}-\bar{\textbf{Y}}_{k,n,J})^\top$.

The tail of the following distribution can be approximated by the Fisher $F$-distribution:
\begin{equation*} \frac{c(m-p)}{m(p+1)}{\rm RMD}^2(\textbf{Y}_{k,n},\bar{\textbf{Y}}_{k,n,J})\sim F_{p+1,m-p}
\end{equation*}
where $c$ and $m$ are the parameters calculated by an algorithm of \cite{hardin2005distribution}. Consequently, the outliers are those whose RMD values exceed the 0.993 quantile of $F_{p+1,m-p}$. Under the RMD criteria, the {\rm VO} part contains the variation properties of the curves. However, its importance goes down with the increase in dimension. Overall, the RMD value is a synthesized index for shape and magnitude outliers.

\section{Trajectory Functional Data Visualization Tools}
\label{s3}
\subsection{Wiggliness of Directional Outlyingness}
Recall that trajectory functional data record the traces of movements from a group of objects, so the most interesting and most common differences between the curves come from the variations of their shapes. Thus, we propose a new tool that specifically detects shape outliers from trajectory functional data, and call it {\it wiggliness of directional outlyingness}. Assuming that the outlyingness function is twice differentiable,  we first compute the integral of the squared second-order derivative of directional outlyingness, then use its $L_2$ norm, as follows:
\begin{equation*}
 {\rm WO}(\textbf{X},F_{\textbf{X}})=\int_{\mathcal{I}}  \left\|\textbf{O}''(\textbf{X}(t),F_{\textbf{X}(t)})\right\|_2^2\omega(t){\rm d}t, 
 \end{equation*}
where $\omega(t)$ is a weight function on $\mathcal{I}$, and the $\textbf{O}''(\textbf{X}(t),F_{\textbf{X}(t)})$ is a vector of the second-order derivatives of each component of the directional outlyingness function with respect to time. We choose $\omega(t)$ as a constant weight function in this paper. The existence of second-order derivatives can be guaranteed if the trajectories are smooth, and projection depth or Mahalanobis depth are applied to derive the directional outlyingness. When the trajectories are observed with random errors, we may approximate them with smoothing splines.

It is well accepted that the second-order derivative is often used to describe the ``wiggliness" of functions. In the smoothing spline model, the sum of the square of the second-order derivative is a classical penalty term for the roughness. From this perspective, {\rm WO} is good at capturing the wiggliness behavior, and is therefore an effective way to detect shape outliers, especially for the curves with large shape variability but close to the center.

\subsection{Properties of WO}
We study some properties of WO in the following theorem.\\
\textbf{Theorem 1} (Transformation invariance). 
Let $\textbf{T}(\textbf{X})$ be a functional, having expression $\textbf{T}(\textbf{X})=\textbf{A}(t)\textbf{X}(t)+\textbf{b}(t)$, where $\textbf{A}(t)=f(t)\textbf{A}_0$ with $f(t)>0$ and $\textbf{A}_0$ an orthogonal matrix, and $\textbf{b}(t)$ is a $p$-dimensional vector, for each $t \in \mathcal{I}$.  Then,
\begin{equation*} {\rm WO}(\textbf{T}(\textbf{X}),F_{{\textbf{T}}(\textbf{X})})={\rm WO}(\textbf{X},F_{\textbf{X}}).
\end{equation*}
We provide the proof of Theorem 1 in the Appendix.

In applications, we usually calculate the WO at a finite set of time points; for example, $T_k=\{t_1,t_2,\ldots,t_k\}$ in $\mathcal{I}$, for a finite sample of trajectories. Therefore, we use the following sample version to calculate WO:
\begin{equation*}
   {\rm WO}_{{\rm T}_k}(\textbf{X},F_{\textbf{X}})=\frac{1}{k}\sum_{i=1}^k \left\|\textbf{O}''(\textbf{X}(t_i),F_{\textbf{X}(t_i)})\right\|_2^2\omega(t_i),
\end{equation*}
where $\textbf{O}''(\textbf{X}(t_i),F_{\textbf{X}(t_i)})$ are approximated by an order-2 difference, at $t=t_i$.

Next, we study the distribution of ${\rm WO}$ when $\textbf{X}$ is generated from a Gaussian random process, which is the most common case. We assume that $\textbf{X}(t)=\{\textbf{X}_1(t),\textbf{X}_2(t)\}^\top$ is generated from a bivariate stationary Gaussian process with zero mean and a Mat\'{e}rn cross-covariance function \citep{gneiting2010matern,apanasovich2012valid}, $$C_{ij}(s,t)=\rho_{ij}\sigma_i\sigma_j\mathcal{M}(|s-t|;\nu_{ij},\alpha_{ij}),~ i,j=1,2,$$ where $\mathcal{M}$ denotes the Mat\'{e}rn class of correlation functions \citep{maternspatial}. We choose $\sigma_1=\sigma_2=1$, $\alpha_{11}=0.02$, $\alpha_{22}=0.01$, $\alpha_{12}=0.016$, $\nu_{11}=1.2$, $\nu_{22}=0.6$, $\nu_{12}=1$, $\rho_{12}=0.6$ and generate two groups of 5000, 10000 samples with $k=1000$ time points.

\begin{figure}[t!]
\centering
\includegraphics[height=10cm,width=10cm]{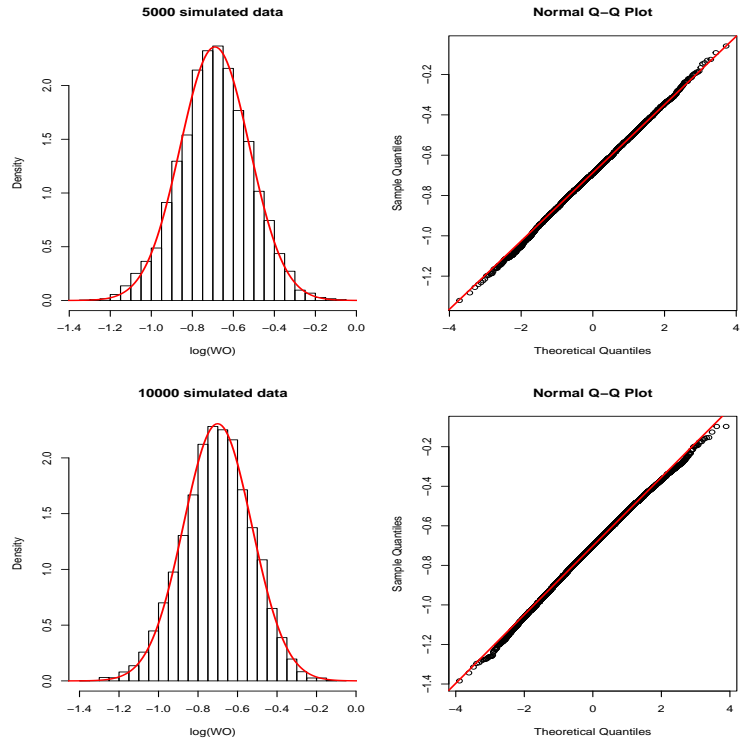}
\caption{Left: histogram of $\log({\rm WO})$ for the simulated data from the  bivariate stationary Gaussian process described below, the red curve stands for a normal distribution with same mean and variance as the histogram; right:  Q-Q plot of the $\log({\rm WO})$.}
\label{2}
\end{figure}

We calculate the ${\rm WO}$ and apply the log transformation. 
The distribution of ${\rm log(WO)}$ can be approximated by a normal distribution, as shown in Figure {\color{blue}{\ref{2}}}.
After normalizing the resulting values, we can approximate the cutoff value by a Gaussian quantile. For example, we can view the $i$-th sample as a potential outlier, if 
\begin{equation}\frac{\log({\rm WO})_i-{\rm med}\{\log({\rm WO})\}}{{\rm MAD}\{\log({\rm WO})\}}>\Phi^{-1}(\alpha), \label{rule}
\end{equation}
where $\Phi(\cdot)$ denotes the standard normal cumulative distribution function, ${\rm med}(\cdot)$ denotes the median, and ${\rm MAD}(\cdot)$ denotes the median absolute deviation. Thus, the cutoff value for outliers can be set by controlling $\alpha$, and we can vary $\alpha$ under different situations to visualize the changes of the flagged outliers. A commonly used value for $\alpha$ is 0.975. This method focuses mainly on the detection of the outliers and, as shown in Section~4, is not suitable for constructing a ranking of the curves that exhibits a reasonable geometric structure. 

\subsection{Trajectory Functional Boxplots}
\begin{figure}[t!]
\centering
\includegraphics[width=17.3cm,height=6cm]{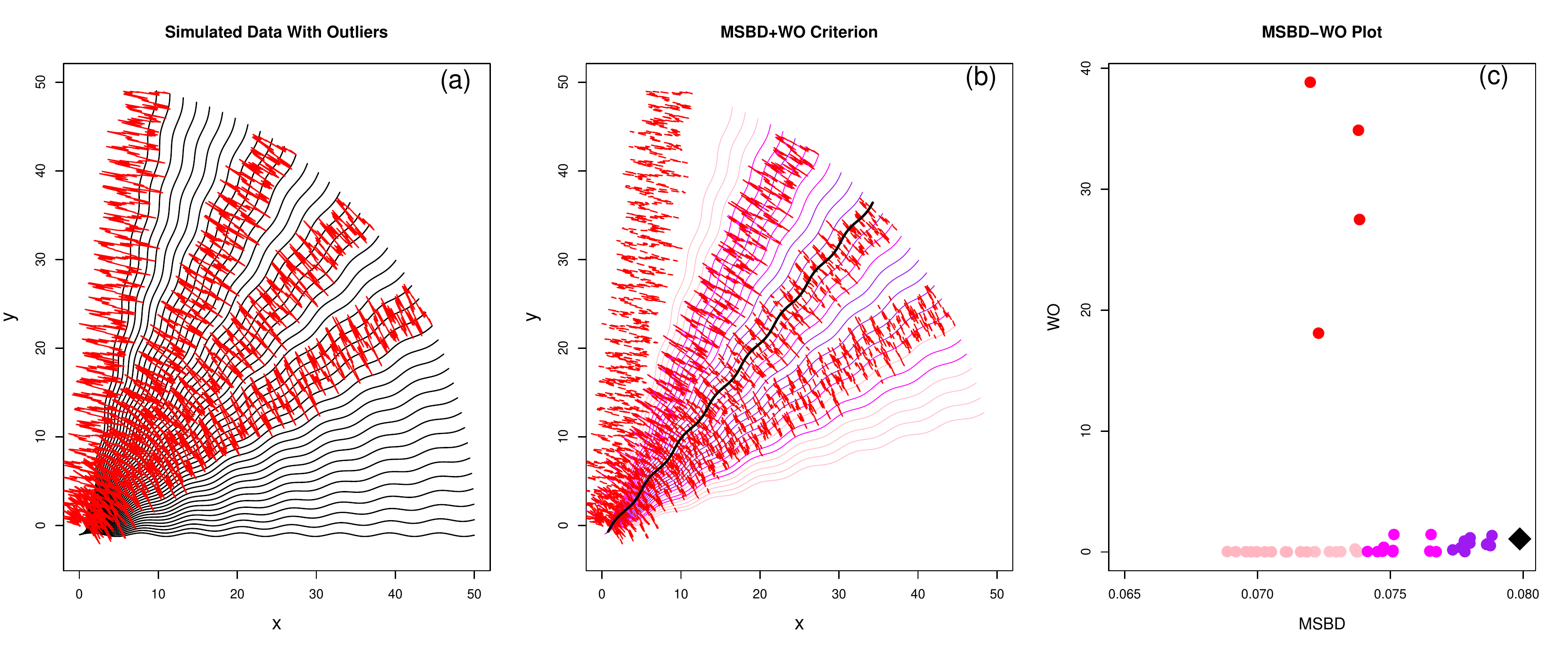}
\caption{\label{fig:frog} (a) Simulated trajectory functional data with outliers denoted by red curves. (b) Trajectory functional boxplot, where black, purple, magenta, pink and red curves represent the median, 25\%, 50\%, 75\% bands and outliers, respectively. (c) MSBD-WO plot. Black, purple, magenta, pink and red points represent the median, 25\%, 50\%, 75\% bands and outliers, respectively.}
\label{3}
\end{figure}

We first construct a box-type plot for trajectory functional data, named {\it trajectory functional boxplot}, that visualizes different levels of central regions, as well as the outliers. Concretely, the trajectory functional boxplot is constructed through the following procedure:
\begin{itemize}
\item [1.] Detecting outliers using criterion (\ref{rule}) and setting the outliers aside from the dataset;
\item [2.] Ranking the remaining data with MSBD to get the center-outward ordering;
\item [3.] Plotting the median and bands formed by a specific percentage (e.g., 25\%, 50\%, 75\%) of data with different colors, and then adding the outliers back to the plot.
\end{itemize}
We provide one example of trajectory functional boxplot in Figure {\color{blue}{\ref{3}}}.  The raw data in  Figure {\color{blue}{\ref{3}}(a)} are generated from Model 2 in the simulation study, where we introduce four shape outliers (the red curves). Figure {\color{blue}{\ref{3}}(b)} shows the trajectory functional boxplot constructed following the above procedure. The outliers detected by WO with $\alpha=0.975$ are presented as dashed red curves, the median curve is the solid black curve; the different levels of central regions, derived by MSBD, are in purple ($25\%$), magenta ($50\%$), and pink ($75\%$) colors. The combination of WO and MSBD makes the trajectory functional boxplot advantageous for both the construction of central regions and the detection of shape outliers.

\subsection{MSBD-WO Plot}

Another tool proposed in this paper is the MSBD-WO plot, which is a scatterplot of points ${\rm (MSBD,~WO)}$, as shown in Figure {\color{blue}{\ref{3}(c)}}. This scatterplot can be used to visualize the distribution of MSBD and WO values for each curve. 
We expect the most central curve with little shape variability to lie in the bottom-right region of the graph (small WO and large MSBD). The central curves with a large shape variability are mapped to the upper-right region (large WO and large MSBD). The outlying curves with a large shape variability correspond to the upper-left region (large WO and small MSBD), and the outlying curves with a small shape variability correspond to the lower-left region (small WO and small MSBD). 
Another possibility would be to combine RMD and MSBD. However, MSBD-RMD plots would not be able to distinguish shape outliers and magnitude outliers as well as MSBD-WO plots because a small MSBD would lead to a large RMD.

\section{Simulation Studies} 
\label{s4}

To assess the effectiveness of our method for the detection of outliers, we conduct a series of simulation studies. We also compare our method with other outlier detection methods described in Section \ref{s3}. 
To investigate the performance of an outlier detector, we use two common measures: $p_c$, the true positive rate (the number of correctly detected outliers divided by the total number of outlying curves), and $p_f$, the false positive rate (the number of falsely detected outliers divided by the number of non-outlying curves). We consider the following four models of trajectories with various shapes and types of contamination.

\subsection{Simulation Design}
\textbf{Model 1}: Shape outliers with small variations 
~\\
The main body includes 70 lines with different slopes, as follows:
\begin{equation*}
\begin{aligned}
&Y_i(t)=k_it+e(t), \quad  e(t) \sim \mathcal{N}(0,1), \quad t \in (0,100), \quad i=1,\ldots,70,  \\
&k_i=\tan \theta_i, \qquad \theta_i=1^{\circ}, 2^{\circ},\ldots, 70^{\circ}.
\end{aligned}
\end{equation*}

\begin{figure}[t!]
\centering
\includegraphics[width=17.3cm,height=12.5cm]{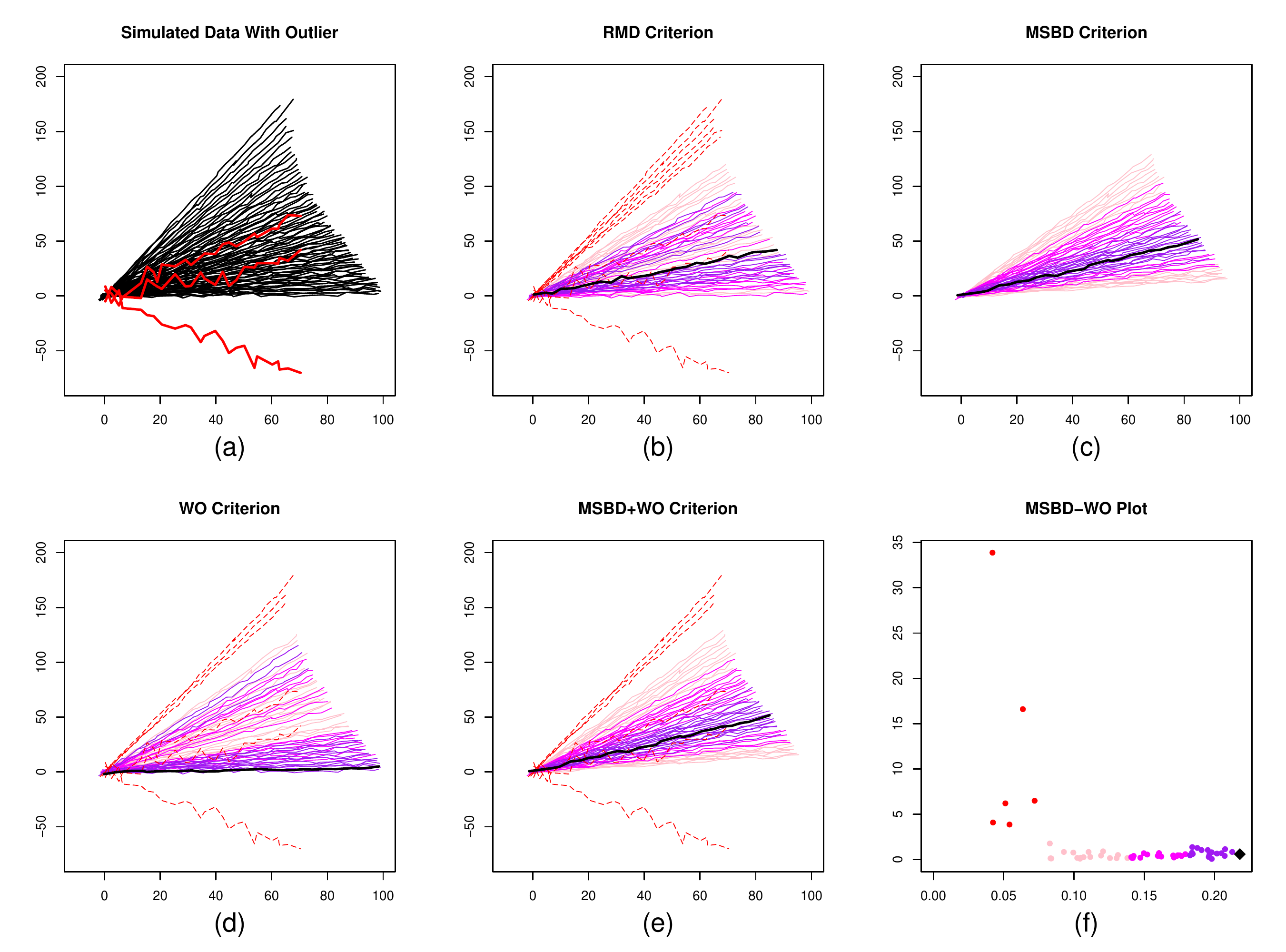}
\caption{\label{fig:frog} Model 1: (a) The generated trajectories with different outliers. (b), (c), (d), (e) and (f) The ranking results by RMD, MSBD and WO, where the black, purple, magenta, pink and red curves (or points) stand for the median, 25\%, 50\%, 75\% bands and outliers, respectively.}
\label{4}
\end{figure}

We add three contaminated outliers, with the first two near the center with larger variations (shape outliers). The third outlier is far from the center, and exhibits the same variations as the first two (outlying for both shape and magnitude):
\begin{equation*}
\begin{aligned}
& Y_1(t)=t+\xi(t),   \qquad Y_2(t)=0.5t+\xi(t), \qquad Y_3(t)=-t+\xi(t), \qquad \xi(t)\sim \mathcal{N}(0,6).
\end{aligned}
\end{equation*}
An example of trajectories from Model 1 is presented in Figure {\color{blue}{\ref{4}(a)}}.

\textbf{Model 2}: Shape outliers with large variation

We generate a sinusoid function and rotate it through the following rotation matrix:
\begin{equation*}
\begin{aligned}
& y(t)=\sin\{x(t)\}, \qquad
\begin{pmatrix}
y_i(t) \\
x_i(t)  
\end{pmatrix}=\begin{pmatrix}
\cos\theta_i & -\sin\theta_i\\ 
\sin \theta_i &  \cos \theta_i 
\end{pmatrix}  \begin{pmatrix}
y(t) \\
x(t) 
\end{pmatrix}, \qquad \theta_i=2^{\circ}, 4^{\circ},\ldots ,80^{\circ}.
\end{aligned}
\end{equation*}
We add four outliers with $y(t)=2\sin\{4x(t)\} +\epsilon(t)$ , and rotate them by $\theta_i=30^{\circ}, 45^{\circ}, 60^{\circ}, 80^{\circ}$, where $\epsilon(t) \sim \mathcal{N}(0,2)$. An example of trajectories from Model 2 is presented in Figure {\color{blue}{\ref{5}(a)}}.

\noindent \textbf{Model 3}: Classical closed-shape outliers

We generate a series of circles with increasing radius and noise:
\begin{equation*}
\begin{aligned}
& x_{i,j}=r_{i}\cos \theta_j+\epsilon_{i,j},\qquad y_{i,j}=r_{i}\sin \theta_j+\epsilon_{i,j},
\end{aligned}
\end{equation*}
where $r_{i\cdot}=20+8*i$ for $i=1,\dots, 20$.
The noises $\epsilon_{i,j}$ are generated from a standard normal distribution, and $\theta_j=j^{\circ}$ for $j=1,\dots, 360$.
The contaminations include one circle with larger noises and three ellipses with the same level of noise as the non-outlying curves. 
An example of trajectories from Model 3 is presented in Figure {\color{blue}{\ref{6}(a)}}.

\noindent\textbf{Model 4}: Special closed-shape outliers

This model has the same main body as Model 3, but is contaminated differently. Specifically, we add rose curves with different leaves as outlying observations. An example of trajectories from Model 4 is presented in Figure {\color{blue}{\ref{7}(a)}}.

We run the simulation with 1000 replications and evaluate the empirical ${p}_c$, ${p}_f$ and their standard deviations. A good performance is usually defined as a high correct detection percentage $p_c$, and a low false detection percentage $p_f$. For the simplicial band depth criterion, the constant factor $F^*=1.5$ is based on a previous simulation study by \cite{sun2011functional,sun2012adjusted}. We set the cutoff value through $c$ and $m$ by the algorithm of \cite{hardin2005distribution} in the RMD criterion; we choose $\alpha=0.975$ as the cutoff values for the detection of outliers in the WO criterion.
\vspace{-.18cm}

\subsection{Outlier Detection and Visualization}
\vspace{-.1cm}

\begin{figure}[t!]
\centering\includegraphics[width=17.3cm,height=12.5cm]{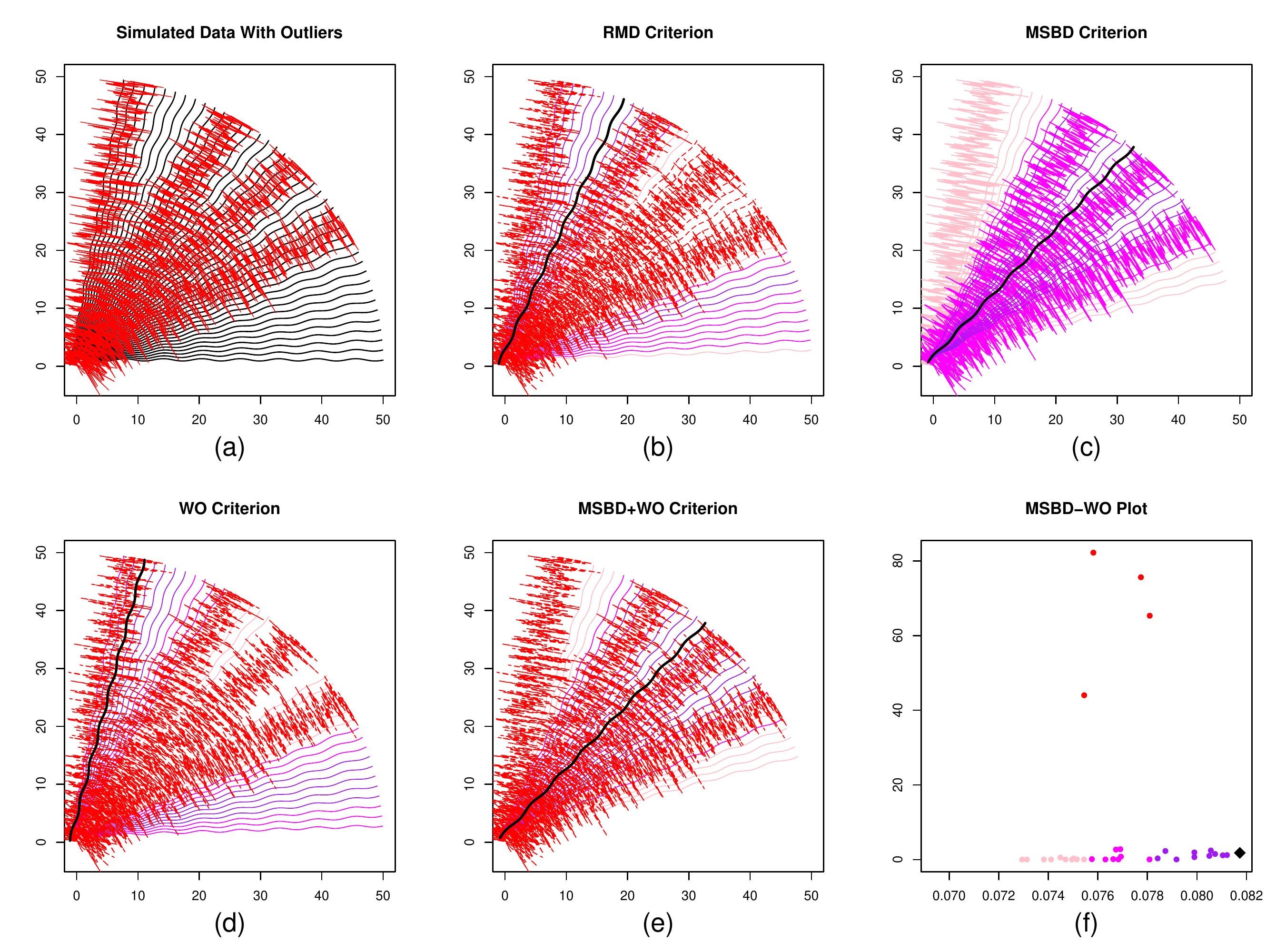}
\caption{\label{fig:frog} Model 2: (a) Generated trajectories with different outliers. (b), (c), (d), (e) and (f) Ranking results by RMD, MSBD and WO, where black, purple, magenta, pink and red curves (or points) represent the median, 25\%, 50\%, 75\% bands and outliers, respectively.}
\label{5}
\end{figure}

\begin{figure}[t!]
\centering
\includegraphics[width=17.3cm,height=12.5cm]{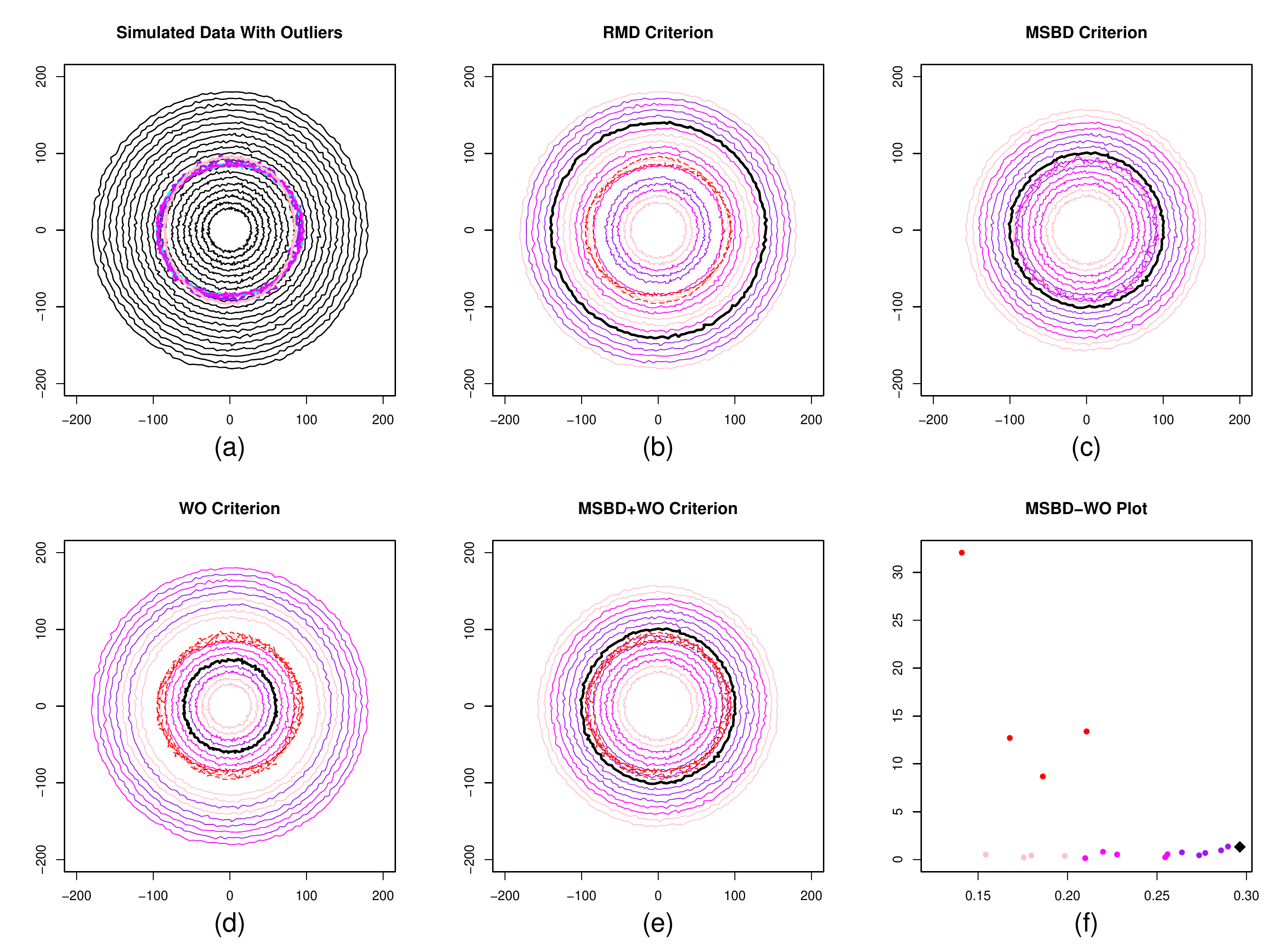}
\caption{\label{fig:frog} Model 3: (a) Generated closed curves with different outliers. (b), (c), (d), (e) and (f) Ranking results by RMD, MSBD and WO, where black, purple, magenta, pink and red curves (or points) represent the median, 25\%, 50\%, 75\% bands and outliers, respectively.}
\label{6}
\end{figure}

\begin{figure}[t!]
\centering
\includegraphics[width=17.3cm,height=12.5cm]{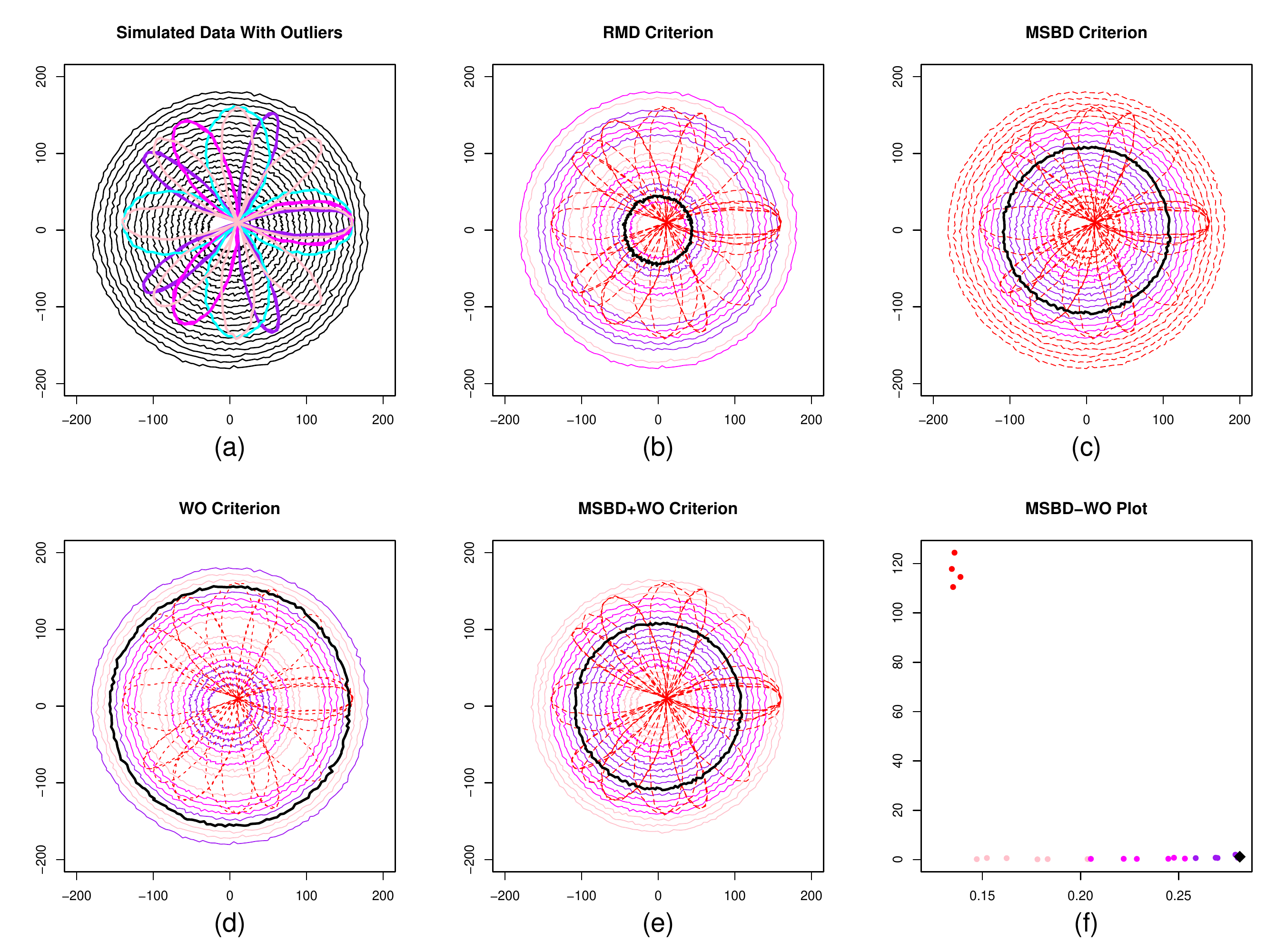}
\caption{\label{fig:frog} Model 4: (a) Generated closed curves with different outliers. (b), (c), (d), (e) and (f) Ranking results by RMD, MSBD and WO, where the black, purple, magenta, pink and red curves (or points) represent the median, 25\%, 50\%, 75\% bands and outliers, respectively.}
\label{7}
\end{figure}

In general, after ranking the data by different criteria, we choose the most central 25\% curves, 25\%-50\% curves, 50\%-75\% curves as our 25\%, 50\% and 75\% bands, respectively. The outliers under different criteria are defined in Section \ref{s3}. Figures \ref{4}-\ref{7} show the plots generated by different criteria.
As we can see from Figure {\color{blue}{\ref{4}}}, MSBD gives a reasonable ranking sequence, from inside to outside. However, the shape outliers in the middle are not easy to detect and they lead to a low $p_c$. On the other hand, it is less likely to have some falsely detected curves in Model 1. In the RMD case, it does well in discovering all the shape outliers and provide a high $p_c$. Nevertheless, it shows a higher false detection rate because it considers both the magnitude and shape parts of the abnormality.

It is worth noting that, for RMD, the ranking results for the 50\% and 75\% bands seem chaotic and irregular, and do not provide a good ranking sequence for constructing a boxplot. For WO, the performance on the detection of shape outliers is excellent, as it shows a high $p_c$ and a low $p_f$. However, the ranking sequence, in this case, is also a disorder. Therefore, it is inappropriate to construct the body part of the boxplots using WO. 

Overall, RMD combines the shape and magnitude behaviors of curves, but MSBD and WO, in this simple case, are more advantageous for ranking sequences and detecting shape outliers, respectively. In Model 1, we demonstrate that our WO criterion has a good performance in detecting shape outliers among the simple straight lines. The patterns in the first five sub-plots of Figure \ref{4} are slightly different because we did not show the 75\%-100\% band for each detection method; this also applies to Figures \ref{5}-\ref{7}.

Concerning the MSBD-WO plots, the properties of magnitude and shape variability for each curve can be seen on the x-axis and y-axis, respectively. From left to right, the depth value increases with the curves, moving from outside to the center. The black rhombus point has the largest depth value, and therefore stands for the median. From bottom-up, the curves show more and more shape variation and are more likely to be detected as shape outliers.  

In Model 2, we find that, under comparatively large variations (the sinusoid curves versus straight lines with variations), the shape outliers detection procedures still perform well for RMD and WO, but that the drawbacks are still that ranking results for the curves do not give a sequence from the center to outside. The 50\% and 75\% bands reverse their sequence in both the RMD and WO criteria. RMD shows a higher false detection rate (Figure \ref{5} or Table \ref{t2}), whereas WO shows a fairly good false detection rate. Their medians also seem unreasonable. The advantage for MSBD remains that it provides a reasonable ranking sequence; however, it has a very high false detection rate and many non-outlying curves are detected as outliers. Therefore, combining the advantage of MSBD and WO gives us a good performance in both ranking sequence and shape outlier detections, resulting in the trajectory functional boxplot shown in Figure {\color{blue}{\ref{5}(e)}}.
The simulation study gives similar results, and shows the robustness of WO in detecting the shape outliers with higher variability. These open straight curves have many applications in migration paths.

Besides the open curves discussed in Models 1 and 2, we investigate the performance of these methods for closed curves. Closed curves have many real applications in medical diagnosis (e.g., vascular malformation).
We test the performance of the outlier detection criterion for closed curves in Models 3 and 4. As we can see from Figures {\color{blue}{\ref{6}}}, {\color{blue}{\ref{7}}} and Table {\color{blue}{\ref{t2}}}, the outlier detection results of WO are still good under these closed curves circumstances; this indicates the robustness of our method for the detection of shape outliers. MSBD also acts well in ranking the functional data, and gives a favorable ranking sequence. Also, it provides a reasonable median curve, compared to WO and RMD, but it shows an unsatisfying classification for the non-outlying curves. RMD's performance is similar to that for the first two models.

\begin{table}[b!]
\caption{Models 1-4: Comparison results among RMD, MSBD and WO in simulation studies for 1000 replicates. $SD$ denotes standard deviations over the replicates.
}
\centering  
\begin{tabular}{lcccclcccc}  

\hline
 Model 1 &${p}_c$ &$SD({p}_c)$ & ${p}_f$ & $SD({p}_f)$ &Model 2 &${p}_c$ &$SD({p}_c)$ & ${p}_f$ & $SD({p}_f)$\\ 
\hline  
RMD&1 &0 &0.11  &0.003 & &1 &0 &0.28  &0.003\\        
MSBD &0.67  &0.02 &0 &0 & &0.25 &0.02 &0 &0\\
WO &1 &0 &0 &0 & &1 &0 &0.12 &0.006\\

\hline

Model 3 &${p}_c$ &$SD({p}_c)$ & ${p}_f$ & $SD({p}_f)$ &Model 4 &${p}_c$ &$SD({p}_c)$ & ${p}_f$ & $SD({p}_f)$\\ 
\hline  
RMD &1 &0 &0  &0 & &1 &0 &0  &0\\         
MSBD &0.75 &0.04 &0.4 &0 & &1 &0 &0.4 &0\\
WO &1 &0 &0 &0 & &1 &0 &0 &0\\
 \hline
\end{tabular}

\label{t2}
\end{table}

Overall, the WO shows its strength in detecting the shape outliers, whereas MSBD can always give a better ranking sequence. In principle, this phenomenon is understandable because MSBD defines outliers as the curves exceeding a certain threshold distance from the center, but it considers the shape variabilities less. Thus, it is reasonable to combine the strengths of both criteria to build our trajectory functional boxplots.

In Table \ref{t2}, we report the simulation results based on 1000 replicates comparing three methods in detecting shape outliers. It shows the good performance of WO with a high $p_c$ value and a low $p_f$ value, and with small standard deviations. RMD gives good results for the detection of outliers, but its $p_f$ value is high and with large standard deviations in Models 1 and 2. MSBD, in many cases, does not have good results for $p_c$ and $p_f$.
In practice, users can modify the $\alpha$ value to see the changes in the detected outliers.

\section{Data Applications} \label{s5}

Besides simulation studies, we examine the two visualization tools, the trajectory functional boxplot and the MSBD-WO plot, on three datasets. Our datasets contain open-straight trajectory functional data and mixtures of open and closed trajectories.

\subsection{Hurricane Paths}
The first dataset consists of hurricane paths. The whole dataset contains 1000 trajectories of longitude and latitude recorded along five common time points in the Caribbean Sea. Because the hurricane path predictions are of interest to many researchers, \cite{cox2013visualizing} established an algorithm for generating an ensemble of hurricane paths, based on historical data. \cite{mirzargar2014curve} constructed a curve boxplot using 50 hurricane tracks simulated with the same algorithm. The raw trajectories are shown in Figure~{\color{blue}{\ref{1}(a)}}. It is evident that direct visualization gives more information about the uncertainty of the hurricane path. A hurricane path can be seen as bivariate functional data, for which the explanatory variable is time, and the two response variables are the longitude and latitude.

We apply the two visualization tools to assess the centrality of hurricane paths and set a series of values for $\alpha$ ranging between 0.9 to 0.99; a visuanimation \citep{Gentonetal2015} of the results is presented in Movie {\color{blue}{\ref{9}}}. The black curve represents the median ranked by MSBD, which is the rightmost point in the MSBD-WO plot. Purple curves represent the 25\% band, magenta curves represent the 50\% band, and pink curves represent the 75\% band. We can find their ranking sequence in the MSBD-WO plot. The red curves are the outliers based on the WO criteria.

\renewcommand{\figurename}{Movie}
\setcounter{figure}{0}
\begin{figure}[b!]
\centering
\animategraphics[width=\textwidth,autoplay,loop,controls]{1}{huu_}{0}{9}
\caption{\label{fig:frog}Left: trajectory functional boxplot for hurricane paths. Right: MSBD-WO plot. Black, purple, magenta, pink and red dash curves (or points) represent the median, 25\%, 50\%, 75\% bands and outliers, respectively. We can see the changes with different $\alpha$ values.}
\label{9}
\end{figure}
\renewcommand{\figurename}{Figure}
\setcounter{figure}{7}

Some of the red curves lying in the 50\% central region are detected as outliers due to their shape variability. 
The curves from different central regions exhibit some differences in length, but the more obvious difference is the width of their spread. The outlying curves behave more wildly than the central ones, which makes these curves become longer.
Overall, the above findings are consistent with our conclusions in simulation studies of the Model 1 for open-straight trajectories. Movie \ref{9} shows the change of different shape outlier results with the change of the $\alpha$ value from 0.9 to 0.99. The recommended $\alpha$ value in the real applications is 0.975, as discussed in the simulation studies, but users have the flexibility to change it.

The trajectory functional boxplot is a valuable tool to visualize hurricane paths, and to give warning to people living nearby. People who live in the 50\% central region may experience severe damage due to hurricanes. Therefore, it is sensible to evacuate the population before the landing of the hurricane. People also receive more information about possible outlying paths. Those who live in Texas may experience the effects of dangerous hurricanes, even if they are not covered by the 50\% and 75\% central regions. The central outlying trajectories show that there is a significant probability that a hurricane may turn Westward, even if it has already landed in Alabama.

\subsection{Migration Patterns}
We consider ecological applications to two datasets of migration patterns: the Tsinghua waterfowl data and the petrel distribution data.

The Tsinghua waterfowl dataset is from Movebank (\citeauthor{si2018spring}, \citeyear{si2018spring}). It contains Spring migration patterns, habitat use and stop-over site protection status for two declining waterfowl species wintering in China, as revealed by satellite tracking. It has GPS information about the routes of the waterfowl. In this case, the paths are complex. Some waterfowls may stay at someplace for a few days (which means their paths have different lengths). Some have a round-trip (viewed as closed curves), and some have straight trajectories (viewed as open curves).

\begin{figure}[b!]
\centering
\includegraphics[width=1\textwidth]{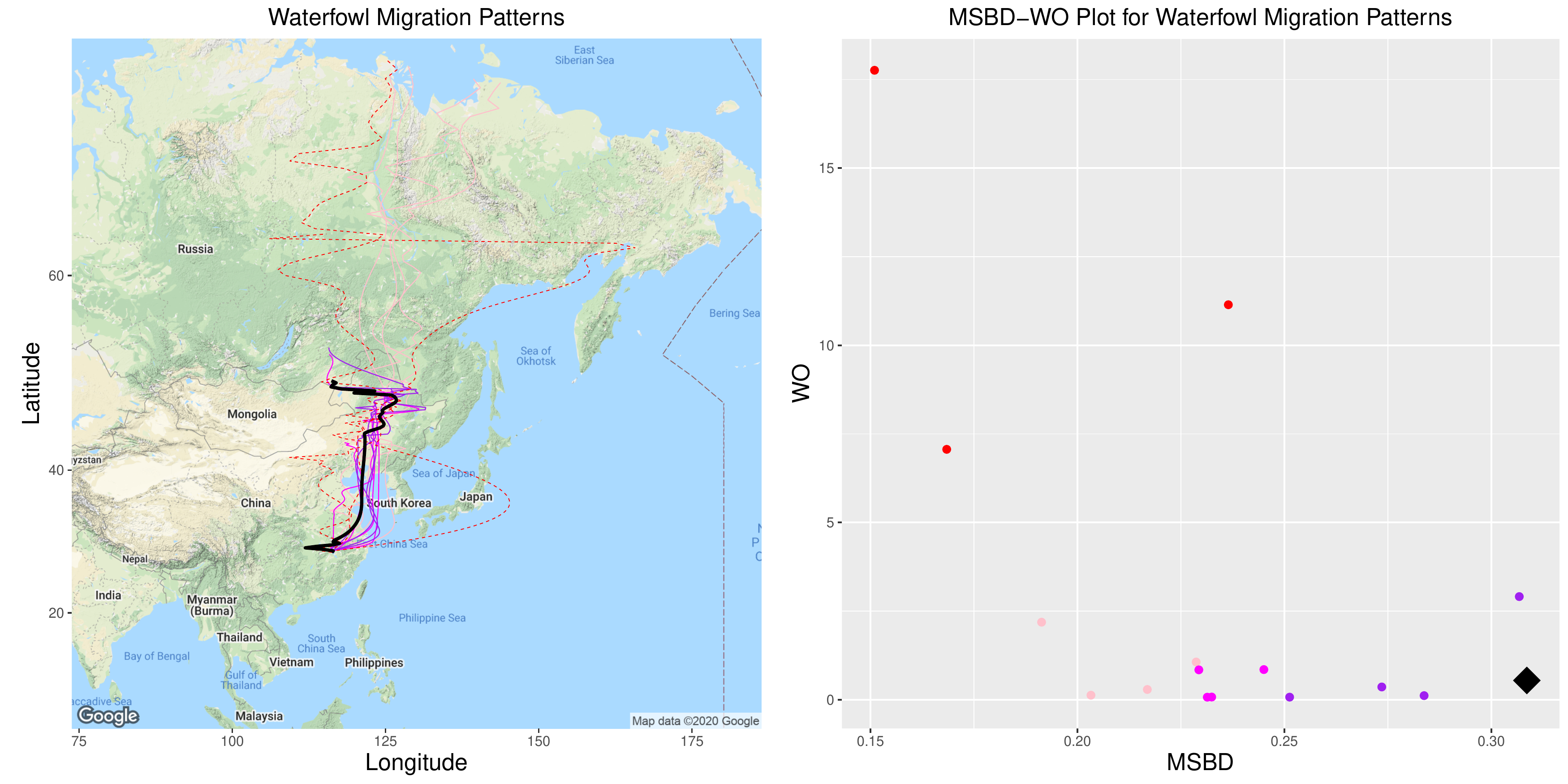}
\caption{\label{fig:frog}Left: Trajectory functional boxplot for waterfowl migration patterns. Right: The MSBD-WO plot. The black, purple, magenta, pink and red dash curves (or points) stand for the median, 25\%, 50\%, 75\% bands and outliers, respectively.}
\label{10}
\end{figure}
 \begin{figure}[!t]
\centering
\includegraphics[width=1\textwidth]{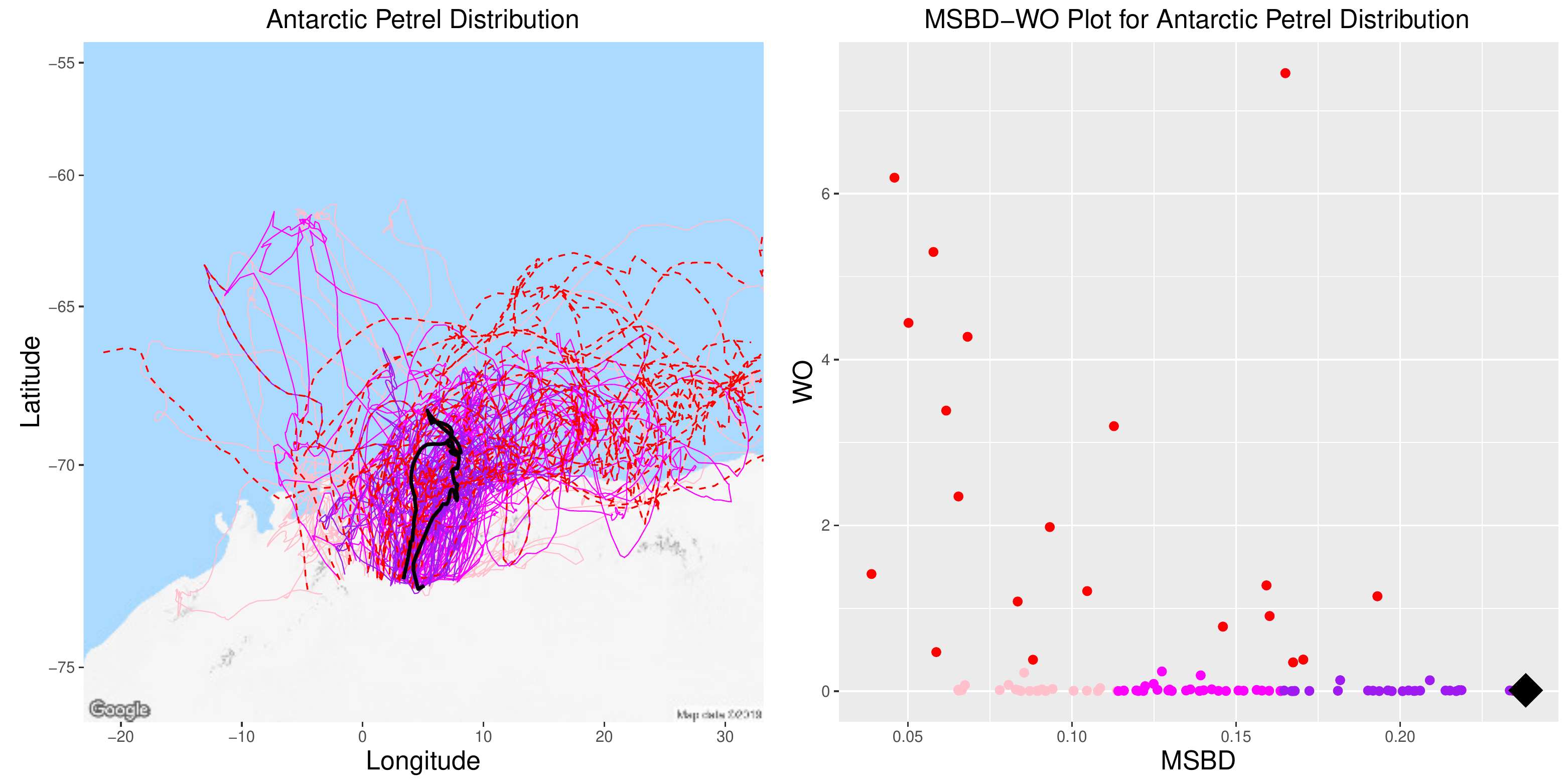}
\caption{\label{fig:frog}Left: Trajectory functional boxplot for petrel distribution. Right: MSBD-WO plot for petrel distribution. Black, purple, magenta, pink and red dash curves (or points) represent the median, 25\%, 50\%, 75\% bands and outliers, respectively. }
\label{11}
\end{figure}

In this study, we view all the routes as bivariate functional data along time.
After some necessary cleaning of the data, we choose 24 bird migration trajectories as our raw data. Because the recording frequencies are different, some of them have 1000 time-point records for longitude and latitude, whereas others only have 200 points. Therefore, we use a cubic smoothing spline to fit different trajectories, and choose 200 common time points for all 24 birds. The raw trajectories are shown in Figure~{\color{blue}{\ref{1}(b)}}. Further, we align these curves so that they start from the same spot.
With a cutoff value $\alpha=0.975$, we obtain the trajectory functional boxplots and the corresponding MSBD-WO plot  shown in Figure {\color{blue}{\ref{10}}}.

The trajectory functional boxplot gives us a meaningful representation of routes of waterfowl migration that provides more information to study and observe their behavior from an ecology perspective. Specifically, we can build more stations in the region covered by the 50\% band to record the migration pattern for the birds. The weird outlier migration paths might occur due to bad weather or a natural disaster. Based on these results, the biologists may take a further step to investigate their behaviors, according to the different categories.

The second dataset comes from \cite{descamps2016sea}, who studied the impact of an extensive fishery for Antarctic krill Euphausia Superba on marine ecosystems, more specifically, the influence of fishing on petrel, which is a predator of Antarctic krill. 
The study involved recording not only the positions where predators are breeding near the fishing grounds but also those where they are breeding far away. Positions of the birds during the non-breeding season are also included. This dataset involves complex and irregular trajectories depicted in Figure~{\color{blue}{\ref{1}(c)}}. Figure {\color{blue}{\ref{11}}} shows the trajectory functional boxplot and the MSBD-WO plot constructed from these data.

In the data preprocessing part, we apply the same data-cleaning and smoothing procedures as above and choose 124 paths as our processed data. However, in this case, the trajectories are more irregular, some are twisted curves, and some are closed curves, which poses significant challenges to our method. We also set $\alpha=0.975$ in this case.

Similarly to the simulation study of Models 3 and 4, we find that our trajectory functional boxplot detects the shape outliers well; they reveal large variations but located within the central regions, as shown by the red curves in Figure {\color{blue}{\ref{11}}}.
The magenta 50\% band contains the routes where petrels fly not far away from the continent and the pink 75\% band includes the routes where petrels fly either very far away or close to the origin. Outliers are straightforward to view in our trajectory functional boxplot. We need to pay more attention to those central outliers because their routes seem quite irregular and twisted. It appears that the fishery industry has a more significant influence on these petrels. Overall, our method serves as a way to separate different flying patterns. The trajectory  functional boxplot is helpful for studying the behavior patterns of the petrels according to their assigned categories in the plot.


\section{Conclusion} \label{s6}
We introduced two novel exploratory tools, the trajectory functional boxplot and the MSBD-WO plot, for visualizing the centrality and detecting outliers of trajectory functional data. To detect abnormal observations, we proposed a criterion focusing on shape outliers; the MSBD provides a ranking revealing a nested structure that provides a more informative and robust description for the bulk of the data. 
The practical performance of the tools was assessed using hurricane paths, waterfowl migrations, and petrel distributions datasets. 

Trajectory functional data may have covariates as well, for example, the wind speed of the hurricane. These covariates can be included in the ranking based on directional outlyingness for multivariate functional data. Moreover, various data transformations can be considered to improve the rankings further, as investigated by \cite{2018arXiv180805414D}.

\section*{Acknowledgements} \label{TGHdisc}
\quad We thank Dr. Donald H. House and his group at Clemson University for sharing the ensemble hurricane generator code. The research reported in this paper was supported by King Abdullah University of Science and Technology (KAUST).

\baselineskip=19pt

\bibliography{reference}

\begin{thebibliography}{}

\bibitem [\protect \citeauthoryear {%
Apanasovich%
, Genton%
\BCBL {}\ \BBA {} Sun%
}{%
Apanasovich%
\ \protect \BOthers {.}}{%
{\protect \APACyear {2012}}%
}]{%
apanasovich2012valid}
\APACinsertmetastar {%
apanasovich2012valid}%
\begin{APACrefauthors}%
Apanasovich, T\BPBI V.%
, Genton, M\BPBI G.%
\BCBL {}\ \BBA {} Sun, Y.%
\end{APACrefauthors}%
\unskip\
\newblock
\APACrefYearMonthDay{2012}{}{}.
\newblock
{\BBOQ}\APACrefatitle {A valid {M}at{\'e}rn class of cross-covariance functions
  for multivariate random fields with any number of components} {A valid
  {M}at{\'e}rn class of cross-covariance functions for multivariate random
  fields with any number of components}.{\BBCQ}
\newblock
\APACjournalVolNumPages{Journal of the American Statistical
  Association}{107}{497}{180--193}.
\PrintBackRefs{\CurrentBib}

\bibitem [\protect \citeauthoryear {%
Arribas-Gil%
\ \BBA {} Romo%
}{%
Arribas-Gil%
\ \BBA {} Romo%
}{%
{\protect \APACyear {2014}}%
}]{%
arribas2014shape}
\APACinsertmetastar {%
arribas2014shape}%
\begin{APACrefauthors}%
Arribas-Gil, A.%
\BCBT {}\ \BBA {} Romo, J.%
\end{APACrefauthors}%
\unskip\
\newblock
\APACrefYearMonthDay{2014}{}{}.
\newblock
{\BBOQ}\APACrefatitle {Shape outlier detection and visualization for functional
  data: the outliergram} {Shape outlier detection and visualization for
  functional data: the outliergram}.{\BBCQ}
\newblock
\APACjournalVolNumPages{Biostatistics}{15}{4}{603--619}.
\PrintBackRefs{\CurrentBib}

\bibitem [\protect \citeauthoryear {%
Claeskens%
, Hubert%
, Slaets%
\BCBL {}\ \BBA {} Vakili%
}{%
Claeskens%
\ \protect \BOthers {.}}{%
{\protect \APACyear {2014}}%
}]{%
claeskens2014multivariate}
\APACinsertmetastar {%
claeskens2014multivariate}%
\begin{APACrefauthors}%
Claeskens, G.%
, Hubert, M.%
, Slaets, L.%
\BCBL {}\ \BBA {} Vakili, K.%
\end{APACrefauthors}%
\unskip\
\newblock
\APACrefYearMonthDay{2014}{}{}.
\newblock
{\BBOQ}\APACrefatitle {Multivariate functional halfspace depth} {Multivariate
  functional halfspace depth}.{\BBCQ}
\newblock
\APACjournalVolNumPages{Journal of the American Statistical
  Association}{109}{}{411--423}.
\PrintBackRefs{\CurrentBib}

\bibitem [\protect \citeauthoryear {%
Cox%
\ \BBA {} Lindell%
}{%
Cox%
\ \BBA {} Lindell%
}{%
{\protect \APACyear {2013}}%
}]{%
cox2013visualizing}
\APACinsertmetastar {%
cox2013visualizing}%
\begin{APACrefauthors}%
Cox, J.%
\BCBT {}\ \BBA {} Lindell, M.%
\end{APACrefauthors}%
\unskip\
\newblock
\APACrefYearMonthDay{2013}{}{}.
\newblock
{\BBOQ}\APACrefatitle {Visualizing uncertainty in predicted hurricane tracks}
  {Visualizing uncertainty in predicted hurricane tracks}.{\BBCQ}
\newblock
\APACjournalVolNumPages{International Journal for Uncertainty
  Quantification}{3}{2}{143--156}.
\PrintBackRefs{\CurrentBib}

\bibitem [\protect \citeauthoryear {%
Dai%
\ \BBA {} Genton%
}{%
Dai%
\ \BBA {} Genton%
}{%
{\protect \APACyear {2018}}%
}]{%
dai2018multivariate}
\APACinsertmetastar {%
dai2018multivariate}%
\begin{APACrefauthors}%
Dai, W.%
\BCBT {}\ \BBA {} Genton, M\BPBI G.%
\end{APACrefauthors}%
\unskip\
\newblock
\APACrefYearMonthDay{2018}{}{}.
\newblock
{\BBOQ}\APACrefatitle {Multivariate Functional Data Visualization and Outlier
  Detection} {Multivariate functional data visualization and outlier
  detection}.{\BBCQ}
\newblock
\APACjournalVolNumPages{Journal of Computational and Graphical
  Statistics}{27}{}{923-934}.
\PrintBackRefs{\CurrentBib}

\bibitem [\protect \citeauthoryear {%
Dai%
\ \BBA {} Genton%
}{%
Dai%
\ \BBA {} Genton%
}{%
{\protect \APACyear {2019}}%
}]{%
dai2018directional}
\APACinsertmetastar {%
dai2018directional}%
\begin{APACrefauthors}%
Dai, W.%
\BCBT {}\ \BBA {} Genton, M\BPBI G.%
\end{APACrefauthors}%
\unskip\
\newblock
\APACrefYearMonthDay{2019}{}{}.
\newblock
{\BBOQ}\APACrefatitle {Directional outlyingness for multivariate functional
  data} {Directional outlyingness for multivariate functional data}.{\BBCQ}
\newblock
\APACjournalVolNumPages{Computational Statistics \& Data
  Analysis}{131}{}{50--65}.
\PrintBackRefs{\CurrentBib}

\bibitem [\protect \citeauthoryear {%
{Dai}%
, {Mrkvi\v cka}%
, {Sun}%
\BCBL {}\ \BBA {} {Genton}%
}{%
{Dai}%
\ \protect \BOthers {.}}{%
{\protect \APACyear {2018}}%
}]{%
2018arXiv180805414D}
\APACinsertmetastar {%
2018arXiv180805414D}%
\begin{APACrefauthors}%
{Dai}, W.%
, {Mrkvi\v cka}, T.%
, {Sun}, Y.%
\BCBL {}\ \BBA {} {Genton}, M\BPBI G.%
\end{APACrefauthors}%
\unskip\
\newblock
\APACrefYearMonthDay{2018}{}{}.
\newblock
{\BBOQ}\APACrefatitle {{Functional outlier detection and taxonomy by sequential
  transformations}} {{Functional outlier detection and taxonomy by sequential
  transformations}}.{\BBCQ}
\newblock
\APACjournalVolNumPages{arXiv e-prints}{}{}{arXiv:1808.05414}.
\PrintBackRefs{\CurrentBib}

\bibitem [\protect \citeauthoryear {%
Descamps%
\ \protect \BOthers {.}}{%
Descamps%
\ \protect \BOthers {.}}{%
{\protect \APACyear {2016}}%
}]{%
descamps2016sea}
\APACinsertmetastar {%
descamps2016sea}%
\begin{APACrefauthors}%
Descamps, S.%
, Tarroux, A.%
, Cherel, Y.%
, Delord, K.%
, God{\o}, O\BPBI R.%
, Kato, A.%
\BDBL {}others%
\end{APACrefauthors}%
\unskip\
\newblock
\APACrefYearMonthDay{2016}{}{}.
\newblock
{\BBOQ}\APACrefatitle {At-sea distribution and prey selection of Antarctic
  petrels and commercial krill fisheries} {At-sea distribution and prey
  selection of antarctic petrels and commercial krill fisheries}.{\BBCQ}
\newblock
\APACjournalVolNumPages{PloS One}{11}{8}{e0156968}.
\PrintBackRefs{\CurrentBib}

\bibitem [\protect \citeauthoryear {%
Ferraty%
\ \BBA {} Vieu%
}{%
Ferraty%
\ \BBA {} Vieu%
}{%
{\protect \APACyear {2006}}%
}]{%
ferraty2006nonparametric}
\APACinsertmetastar {%
ferraty2006nonparametric}%
\begin{APACrefauthors}%
Ferraty, F.%
\BCBT {}\ \BBA {} Vieu, P.%
\end{APACrefauthors}%
\unskip\
\newblock
\APACrefYear{2006}.
\newblock
\APACrefbtitle {Nonparametric {F}unctional {D}ata {A}nalysis: {T}heory and
  {P}ractice} {Nonparametric {F}unctional {D}ata {A}nalysis: {T}heory and
  {P}ractice}.
\newblock
\APACaddressPublisher{}{Springer}.
\PrintBackRefs{\CurrentBib}

\bibitem [\protect \citeauthoryear {%
Fraiman%
\ \BBA {} Muniz%
}{%
Fraiman%
\ \BBA {} Muniz%
}{%
{\protect \APACyear {2001}}%
}]{%
fraiman2001trimmed}
\APACinsertmetastar {%
fraiman2001trimmed}%
\begin{APACrefauthors}%
Fraiman, R.%
\BCBT {}\ \BBA {} Muniz, G.%
\end{APACrefauthors}%
\unskip\
\newblock
\APACrefYearMonthDay{2001}{}{}.
\newblock
{\BBOQ}\APACrefatitle {Trimmed means for functional data} {Trimmed means for
  functional data}.{\BBCQ}
\newblock
\APACjournalVolNumPages{TEST}{10}{2}{419--440}.
\PrintBackRefs{\CurrentBib}

\bibitem [\protect \citeauthoryear {%
Genton%
\ \protect \BOthers {.}}{%
Genton%
\ \protect \BOthers {.}}{%
{\protect \APACyear {2015}}%
}]{%
Gentonetal2015}
\APACinsertmetastar {%
Gentonetal2015}%
\begin{APACrefauthors}%
Genton, M\BPBI G.%
, Castruccio, S.%
, Crippa, P.%
, Dutta, S.%
, Huser, R.%
, Sun, Y.%
\BCBL {}\ \BBA {} Vettori, S.%
\end{APACrefauthors}%
\unskip\
\newblock
\APACrefYearMonthDay{2015}{}{}.
\newblock
{\BBOQ}\APACrefatitle {Visuanimation in statistics} {Visuanimation in
  statistics}.{\BBCQ}
\newblock
\APACjournalVolNumPages{Stat}{4}{}{81--96}.
\PrintBackRefs{\CurrentBib}

\bibitem [\protect \citeauthoryear {%
Genton%
, Johnson%
, Potter%
, Stenchikov%
\BCBL {}\ \BBA {} Sun%
}{%
Genton%
\ \protect \BOthers {.}}{%
{\protect \APACyear {2014}}%
}]{%
genton2014surface}
\APACinsertmetastar {%
genton2014surface}%
\begin{APACrefauthors}%
Genton, M\BPBI G.%
, Johnson, C.%
, Potter, K.%
, Stenchikov, G.%
\BCBL {}\ \BBA {} Sun, Y.%
\end{APACrefauthors}%
\unskip\
\newblock
\APACrefYearMonthDay{2014}{}{}.
\newblock
{\BBOQ}\APACrefatitle {Surface boxplots} {Surface boxplots}.{\BBCQ}
\newblock
\APACjournalVolNumPages{Stat}{3}{1}{1--11}.
\PrintBackRefs{\CurrentBib}

\bibitem [\protect \citeauthoryear {%
Gneiting%
, Kleiber%
\BCBL {}\ \BBA {} Schlather%
}{%
Gneiting%
\ \protect \BOthers {.}}{%
{\protect \APACyear {2010}}%
}]{%
gneiting2010matern}
\APACinsertmetastar {%
gneiting2010matern}%
\begin{APACrefauthors}%
Gneiting, T.%
, Kleiber, W.%
\BCBL {}\ \BBA {} Schlather, M.%
\end{APACrefauthors}%
\unskip\
\newblock
\APACrefYearMonthDay{2010}{}{}.
\newblock
{\BBOQ}\APACrefatitle {Mat{\'e}rn cross-covariance functions for multivariate
  random fields} {Mat{\'e}rn cross-covariance functions for multivariate random
  fields}.{\BBCQ}
\newblock
\APACjournalVolNumPages{Journal of the American Statistical
  Association}{105}{491}{1167--1177}.
\PrintBackRefs{\CurrentBib}

\bibitem [\protect \citeauthoryear {%
Hardin%
\ \BBA {} Rocke%
}{%
Hardin%
\ \BBA {} Rocke%
}{%
{\protect \APACyear {2005}}%
}]{%
hardin2005distribution}
\APACinsertmetastar {%
hardin2005distribution}%
\begin{APACrefauthors}%
Hardin, J.%
\BCBT {}\ \BBA {} Rocke, D\BPBI M.%
\end{APACrefauthors}%
\unskip\
\newblock
\APACrefYearMonthDay{2005}{}{}.
\newblock
{\BBOQ}\APACrefatitle {The distribution of robust distances} {The distribution
  of robust distances}.{\BBCQ}
\newblock
\APACjournalVolNumPages{Journal of Computational and Graphical
  Statistics}{14}{4}{928--946}.
\PrintBackRefs{\CurrentBib}

\bibitem [\protect \citeauthoryear {%
Horv{\'a}th%
\ \BBA {} Kokoszka%
}{%
Horv{\'a}th%
\ \BBA {} Kokoszka%
}{%
{\protect \APACyear {2012}}%
}]{%
horvath2012inference}
\APACinsertmetastar {%
horvath2012inference}%
\begin{APACrefauthors}%
Horv{\'a}th, L.%
\BCBT {}\ \BBA {} Kokoszka, P.%
\end{APACrefauthors}%
\unskip\
\newblock
\APACrefYear{2012}.
\newblock
\APACrefbtitle {Inference for {F}unctional {D}ata with {A}pplications}
  {Inference for {F}unctional {D}ata with {A}pplications}.
\newblock
\APACaddressPublisher{}{Springer}.
\PrintBackRefs{\CurrentBib}

\bibitem [\protect \citeauthoryear {%
Hubert%
, Rousseeuw%
\BCBL {}\ \BBA {} Segaert%
}{%
Hubert%
\ \protect \BOthers {.}}{%
{\protect \APACyear {2015}}%
}]{%
hubert2015multivariate}
\APACinsertmetastar {%
hubert2015multivariate}%
\begin{APACrefauthors}%
Hubert, M.%
, Rousseeuw, P\BPBI J.%
\BCBL {}\ \BBA {} Segaert, P.%
\end{APACrefauthors}%
\unskip\
\newblock
\APACrefYearMonthDay{2015}{}{}.
\newblock
{\BBOQ}\APACrefatitle {Multivariate functional outlier detection} {Multivariate
  functional outlier detection}.{\BBCQ}
\newblock
\APACjournalVolNumPages{Statistical Methods \& Applications}{24}{}{177--202}.
\PrintBackRefs{\CurrentBib}

\bibitem [\protect \citeauthoryear {%
Hyndman%
\ \BBA {} Shang%
}{%
Hyndman%
\ \BBA {} Shang%
}{%
{\protect \APACyear {2010}}%
}]{%
hyndman2010rainbow}
\APACinsertmetastar {%
hyndman2010rainbow}%
\begin{APACrefauthors}%
Hyndman, R\BPBI J.%
\BCBT {}\ \BBA {} Shang, H\BPBI L.%
\end{APACrefauthors}%
\unskip\
\newblock
\APACrefYearMonthDay{2010}{}{}.
\newblock
{\BBOQ}\APACrefatitle {Rainbow plots, bagplots, and boxplots for functional
  data} {Rainbow plots, bagplots, and boxplots for functional data}.{\BBCQ}
\newblock
\APACjournalVolNumPages{Journal of Computational and Graphical
  Statistics}{19}{1}{29--45}.
\PrintBackRefs{\CurrentBib}

\bibitem [\protect \citeauthoryear {%
Ieva%
\ \BBA {} Paganoni%
}{%
Ieva%
\ \BBA {} Paganoni%
}{%
{\protect \APACyear {2013}}%
}]{%
ieva2013depth}
\APACinsertmetastar {%
ieva2013depth}%
\begin{APACrefauthors}%
Ieva, F.%
\BCBT {}\ \BBA {} Paganoni, A\BPBI M.%
\end{APACrefauthors}%
\unskip\
\newblock
\APACrefYearMonthDay{2013}{}{}.
\newblock
{\BBOQ}\APACrefatitle {Depth measures for multivariate functional data} {Depth
  measures for multivariate functional data}.{\BBCQ}
\newblock
\APACjournalVolNumPages{Communications in Statistics-Theory and
  Methods}{42}{7}{1265--1276}.
\PrintBackRefs{\CurrentBib}

\bibitem [\protect \citeauthoryear {%
L{\'o}pez-Pintado%
, Sun%
, Lin%
\BCBL {}\ \BBA {} Genton%
}{%
L{\'o}pez-Pintado%
\ \protect \BOthers {.}}{%
{\protect \APACyear {2014}}%
}]{%
lopez2014simplicial}
\APACinsertmetastar {%
lopez2014simplicial}%
\begin{APACrefauthors}%
L{\'o}pez-Pintado, S.%
, Sun, Y.%
, Lin, J\BPBI K.%
\BCBL {}\ \BBA {} Genton, M\BPBI G.%
\end{APACrefauthors}%
\unskip\
\newblock
\APACrefYearMonthDay{2014}{}{}.
\newblock
{\BBOQ}\APACrefatitle {Simplicial band depth for multivariate functional data}
  {Simplicial band depth for multivariate functional data}.{\BBCQ}
\newblock
\APACjournalVolNumPages{Advances in Data Analysis and
  Classification}{8}{3}{321--338}.
\PrintBackRefs{\CurrentBib}

\bibitem [\protect \citeauthoryear {%
Mat{\'e}rn%
}{%
Mat{\'e}rn%
}{%
{\protect \APACyear {1960}}%
}]{%
maternspatial}
\APACinsertmetastar {%
maternspatial}%
\begin{APACrefauthors}%
Mat{\'e}rn, B.%
\end{APACrefauthors}%
\unskip\
\newblock
\APACrefYear{1960}.
\newblock
\APACrefbtitle {Spatial {V}ariation} {Spatial {V}ariation}.
\newblock
\APACaddressPublisher{}{Springer}.
\PrintBackRefs{\CurrentBib}

\bibitem [\protect \citeauthoryear {%
Mirzargar%
, Whitaker%
\BCBL {}\ \BBA {} Kirby%
}{%
Mirzargar%
\ \protect \BOthers {.}}{%
{\protect \APACyear {2014}}%
}]{%
mirzargar2014curve}
\APACinsertmetastar {%
mirzargar2014curve}%
\begin{APACrefauthors}%
Mirzargar, M.%
, Whitaker, R\BPBI T.%
\BCBL {}\ \BBA {} Kirby, R\BPBI M.%
\end{APACrefauthors}%
\unskip\
\newblock
\APACrefYearMonthDay{2014}{}{}.
\newblock
{\BBOQ}\APACrefatitle {Curve boxplot: Generalization of boxplot for ensembles
  of curves} {Curve boxplot: Generalization of boxplot for ensembles of
  curves}.{\BBCQ}
\newblock
\APACjournalVolNumPages{IEEE Transactions on Visualization and Computer
  Graphics}{20}{12}{2654--2663}.
\PrintBackRefs{\CurrentBib}

\bibitem [\protect \citeauthoryear {%
Nagy%
, Gijbels%
\BCBL {}\ \BBA {} Hlubinka%
}{%
Nagy%
\ \protect \BOthers {.}}{%
{\protect \APACyear {2017}}%
}]{%
nagy2017depth}
\APACinsertmetastar {%
nagy2017depth}%
\begin{APACrefauthors}%
Nagy, S.%
, Gijbels, I.%
\BCBL {}\ \BBA {} Hlubinka, D.%
\end{APACrefauthors}%
\unskip\
\newblock
\APACrefYearMonthDay{2017}{}{}.
\newblock
{\BBOQ}\APACrefatitle {Depth-based recognition of shape outlying functions}
  {Depth-based recognition of shape outlying functions}.{\BBCQ}
\newblock
\APACjournalVolNumPages{Journal of Computational and Graphical
  Statistics}{26}{}{883--893}.
\PrintBackRefs{\CurrentBib}

\bibitem [\protect \citeauthoryear {%
Ramsay%
\ \BBA {} Silverman%
}{%
Ramsay%
\ \BBA {} Silverman%
}{%
{\protect \APACyear {2005}}%
}]{%
ramsayfunctional}
\APACinsertmetastar {%
ramsayfunctional}%
\begin{APACrefauthors}%
Ramsay, J\BPBI O.%
\BCBT {}\ \BBA {} Silverman, B\BPBI W.%
\end{APACrefauthors}%
\unskip\
\newblock
\APACrefYear{2005}.
\newblock
\APACrefbtitle {Functional {D}ata {A}nalysis {\rm (second ed.)}} {Functional
  {D}ata {A}nalysis {\rm (second ed.)}}.
\newblock
\APACaddressPublisher{}{Springer}.
\PrintBackRefs{\CurrentBib}

\bibitem [\protect \citeauthoryear {%
Rousseeuw%
}{%
Rousseeuw%
}{%
{\protect \APACyear {1985}}%
}]{%
rousseeuw1985multivariate}
\APACinsertmetastar {%
rousseeuw1985multivariate}%
\begin{APACrefauthors}%
Rousseeuw, P\BPBI J.%
\end{APACrefauthors}%
\unskip\
\newblock
\APACrefYearMonthDay{1985}{}{}.
\newblock
{\BBOQ}\APACrefatitle {Multivariate estimation with high breakdown point}
  {Multivariate estimation with high breakdown point}.{\BBCQ}
\newblock
\APACjournalVolNumPages{Mathematical Statistics and
  Applications}{8}{37}{283-297}.
\PrintBackRefs{\CurrentBib}

\bibitem [\protect \citeauthoryear {%
Rousseeuw%
, Raymaekers%
\BCBL {}\ \BBA {} Hubert%
}{%
Rousseeuw%
\ \protect \BOthers {.}}{%
{\protect \APACyear {2018}}%
}]{%
rousseeuw2018measure}
\APACinsertmetastar {%
rousseeuw2018measure}%
\begin{APACrefauthors}%
Rousseeuw, P\BPBI J.%
, Raymaekers, J.%
\BCBL {}\ \BBA {} Hubert, M.%
\end{APACrefauthors}%
\unskip\
\newblock
\APACrefYearMonthDay{2018}{}{}.
\newblock
{\BBOQ}\APACrefatitle {A measure of directional outlyingness with applications
  to image data and video} {A measure of directional outlyingness with
  applications to image data and video}.{\BBCQ}
\newblock
\APACjournalVolNumPages{Journal of Computational and Graphical
  Statistics}{27}{2}{345-359}.
\PrintBackRefs{\CurrentBib}

\bibitem [\protect \citeauthoryear {%
Si%
\ \protect \BOthers {.}}{%
Si%
\ \protect \BOthers {.}}{%
{\protect \APACyear {2018}}%
}]{%
si2018spring}
\APACinsertmetastar {%
si2018spring}%
\begin{APACrefauthors}%
Si, Y.%
, Xu, Y.%
, Xu, F.%
, Li, X.%
, Zhang, W.%
, Wielstra, B.%
\BDBL {}others%
\end{APACrefauthors}%
\unskip\
\newblock
\APACrefYearMonthDay{2018}{}{}.
\newblock
{\BBOQ}\APACrefatitle {Spring migration patterns, habitat use, and stopover
  site protection status for two declining waterfowl species wintering in
  {C}hina as revealed by satellite tracking} {Spring migration patterns,
  habitat use, and stopover site protection status for two declining waterfowl
  species wintering in {C}hina as revealed by satellite tracking}.{\BBCQ}
\newblock
\APACjournalVolNumPages{Ecology and Evolution}{8}{}{6280–6289}.
\PrintBackRefs{\CurrentBib}

\bibitem [\protect \citeauthoryear {%
Sun%
\ \BBA {} Genton%
}{%
Sun%
\ \BBA {} Genton%
}{%
{\protect \APACyear {2011}}%
}]{%
sun2011functional}
\APACinsertmetastar {%
sun2011functional}%
\begin{APACrefauthors}%
Sun, Y.%
\BCBT {}\ \BBA {} Genton, M\BPBI G.%
\end{APACrefauthors}%
\unskip\
\newblock
\APACrefYearMonthDay{2011}{}{}.
\newblock
{\BBOQ}\APACrefatitle {Functional boxplots} {Functional boxplots}.{\BBCQ}
\newblock
\APACjournalVolNumPages{Journal of Computational and Graphical
  Statistics}{20}{2}{316--334}.
\PrintBackRefs{\CurrentBib}

\bibitem [\protect \citeauthoryear {%
Sun%
\ \BBA {} Genton%
}{%
Sun%
\ \BBA {} Genton%
}{%
{\protect \APACyear {2012}}%
}]{%
sun2012adjusted}
\APACinsertmetastar {%
sun2012adjusted}%
\begin{APACrefauthors}%
Sun, Y.%
\BCBT {}\ \BBA {} Genton, M\BPBI G.%
\end{APACrefauthors}%
\unskip\
\newblock
\APACrefYearMonthDay{2012}{}{}.
\newblock
{\BBOQ}\APACrefatitle {Adjusted functional boxplots for spatio-temporal data
  visualization and outlier detection} {Adjusted functional boxplots for
  spatio-temporal data visualization and outlier detection}.{\BBCQ}
\newblock
\APACjournalVolNumPages{Environmetrics}{23}{1}{54--64}.
\PrintBackRefs{\CurrentBib}

\bibitem [\protect \citeauthoryear {%
Wang%
, Chiou%
\BCBL {}\ \BBA {} M{\"u}ller%
}{%
Wang%
\ \protect \BOthers {.}}{%
{\protect \APACyear {2016}}%
}]{%
wang2016functional}
\APACinsertmetastar {%
wang2016functional}%
\begin{APACrefauthors}%
Wang, J\BHBI L.%
, Chiou, J\BHBI M.%
\BCBL {}\ \BBA {} M{\"u}ller, H\BHBI G.%
\end{APACrefauthors}%
\unskip\
\newblock
\APACrefYearMonthDay{2016}{}{}.
\newblock
{\BBOQ}\APACrefatitle {Functional data analysis} {Functional data
  analysis}.{\BBCQ}
\newblock
\APACjournalVolNumPages{Annual Review of Statistics and Its
  Application}{3}{}{257--295}.
\PrintBackRefs{\CurrentBib}

\bibitem [\protect \citeauthoryear {%
Xie%
, Kurtek%
, Bharath%
\BCBL {}\ \BBA {} Sun%
}{%
Xie%
\ \protect \BOthers {.}}{%
{\protect \APACyear {2017}}%
}]{%
xie2017geometric}
\APACinsertmetastar {%
xie2017geometric}%
\begin{APACrefauthors}%
Xie, W.%
, Kurtek, S.%
, Bharath, K.%
\BCBL {}\ \BBA {} Sun, Y.%
\end{APACrefauthors}%
\unskip\
\newblock
\APACrefYearMonthDay{2017}{}{}.
\newblock
{\BBOQ}\APACrefatitle {A geometric approach to visualization of variability in
  functional data} {A geometric approach to visualization of variability in
  functional data}.{\BBCQ}
\newblock
\APACjournalVolNumPages{Journal of the American Statistical
  Association}{112}{519}{979--993}.
\PrintBackRefs{\CurrentBib}

\bibitem [\protect \citeauthoryear {%
Zuo%
\ \BBA {} Serfling%
}{%
Zuo%
\ \BBA {} Serfling%
}{%
{\protect \APACyear {2000}}%
}]{%
zuo2000general}
\APACinsertmetastar {%
zuo2000general}%
\begin{APACrefauthors}%
Zuo, Y.%
\BCBT {}\ \BBA {} Serfling, R.%
\end{APACrefauthors}%
\unskip\
\newblock
\APACrefYearMonthDay{2000}{}{}.
\newblock
{\BBOQ}\APACrefatitle {General notions of statistical depth function} {General
  notions of statistical depth function}.{\BBCQ}
\newblock
\APACjournalVolNumPages{Annals of Statistics}{28}{}{461--482}.
\PrintBackRefs{\CurrentBib}

\end{thebibliography}

\baselineskip=20pt

\section*{Appendix}

{\em Proof of Theorem 1}: \\
According to Theorem 1 of \citet{dai2018directional}, we have the following result for $\textbf{O}(t)$ at a fixed time point: 
$$\textbf{O}(\textbf{T}(\textbf{X}(t)),F_{\textbf{T}(\textbf{X}(t))})=\textbf{A}_0\textbf{O}(\textbf{X}(t),F_{\textbf{X}(t)}).$$
If the function $\textbf{O}(t)$ has a second order derivative then its smoothness is retained through a rotation by the orthogonal matrix $\textbf{A}_0$. Denoting $\textbf{U}(\textbf{X}(t),F_{\textbf{X}(t)})=\textbf{A}_0\textbf{O}(\textbf{X}(t),F_{\textbf{X}(t)})$, it is then obvious that $$\|\textbf{U}''(\textbf{X}(t),F_{\textbf{X}(t)})\|=\|\textbf{O}''(\textbf{X}(t),F_{\textbf{X}(t)})\|,$$ since $\textbf{A}_0$ is orthogonal. For a constant weight function $\omega(t)$, we conclude that 
$${\rm WO}(\textbf{T}(\textbf{X}),F_{\textbf{T}(\textbf{X})})={\rm WO}(\textbf{X},F_{\textbf{X}}).$$
\hfill $\Box$
\end{document}